\documentclass[12pt,preprint]{aastex}

\shorttitle{Interstellar Silicate Dust Toward TXS 0218+357}
\shortauthors{Aller et al.}

\begin{document}

\title{Interstellar Silicate Dust in the z=0.685 Absorber Toward TXS 0218+357}

\author{Monique C. Aller and Varsha P. Kulkarni}
\affil{Department of Physics and Astronomy, University of South Carolina, 712 Main Street, Columbia, SC, 29208, USA}
\email{ALLERM@mailbox.sc.edu}

\author{Donald G. York and Daniel E. Welty}
\affil{Department of Astronomy \& Astrophysics, University of Chicago, 5640 S. Ellis Ave., Chicago, IL, 60637, USA}

\author{Giovanni Vladilo}
\affil{Osservatorio Astronomico di Trieste, Via Tiepolo 11, 34143 Trieste, Italy}

\and

\author{Nicholas Liger}
\affil{University of South Carolina, Columbia, SC, 29208, USA}

\begin{abstract}

We report the detection of interstellar silicate dust in the z$_{abs}$=0.685 absorber along the sightline toward the gravitationally lensed blazar TXS 0218+357. 
Using \textit{Spitzer Space Telescope} \textit{Infrared Spectrograph} data we detect the 10~$\micron$ silicate absorption feature with a detection significance of 10.7$\sigma$.
We fit laboratory-derived silicate dust profile templates obtained from literature to the observed 10~$\micron$ absorption feature, and find that the best single-mineral fit is obtained using an amorphous olivine template with 
a measured peak optical depth of $\tau_{10}$=0.49$\pm$0.02, which rises to $\tau_{10}\sim$0.67$\pm$0.04 if the covering factor is taken into account. 
We also detected the 18~$\micron$ silicate absorption feature in our data with a $>$3$\sigma$ significance. 
Due to the proximity of the 18~$\micron$ absorption feature to the edge of our covered spectral range, and associated uncertainty about the 
shape of the quasar continuum normalization near 18~$\micron$, we do not independently fit this feature. 
We find, however, that the shape and depth of the 18~$\micron$~silicate absorption are well-matched to the amorphous olivine template prediction, given the optical depth inferred for the 10~$\micron$ feature. 
The measured 10~$\micron$ peak optical depth in this absorber is significantly higher than those found in previously studied quasar absorption systems. The reddening, 21-cm absorption, and velocity spread of Mg~II 
are not outliers relative to other studied absorption systems, however. This high optical depth may be evidence for variations in dust grain properties in the ISM between this and the previously studied high redshift galaxies. 
\end{abstract}

\keywords{dust, extinction - galaxies: ISM - quasars: absorption lines}
\section{INTRODUCTION}\label{intro}
The dust and gas that comprise the interstellar medium (ISM) play an important role in galaxy evolution, despite the fact that they constitute a relatively small fraction of each galaxy's total mass (e.g., $\sim$10\% in the Milky Way). 
The presence of interstellar dust is physically manifested both through absorption and scattering processes which redistribute $>$30\% of the starlight emitted at UV-optical wavelengths 
over a range of infrared (IR) wavelengths \citep{Bernstein02}, and through the depletion of refractory elements from ISM gas. 
The physical and chemical states of the ISM are significantly affected by the dust, which regulates the thermal energy balance through the ejection of photoelectrons in photodissociation 
regions, and via IR emission from molecular and shock-heated plasmas. Dust grain surfaces also promote the formation of molecules, such as H$_2$, which act as a coolant facilitating molecular cloud collapse. 
As the ISM gas is formed into new stars and the galaxy's stellar population evolves, the ratio of the silicate dust to the carbonaceous dust reflects the relative abundances of their different stellar sources. 
Silicate dust is typically produced in oxygen-rich environments, such as near Type II supernovae or AGB stars with low C/O ratios, while carbonaceous dust is associated with carbon-rich environments, such as near evolved low-mass stars,
although the details of the formation are still debated both in local and in higher redshift galaxies [e.g., \citet{Dwek11}].  
Since dust is relatively quickly destroyed through shock processing in the ISM, with a typical lifetime of $\approx$500~Myr for many dust species \citep{Jones96,Jones11}, 
interstellar dust is a sensitive probe of the state of the ISM material and stellar populations at a given epoch, assuming that selective destruction and reformation rates in the ISM are not highly variable. 

Quasar absorption systems (QASs) provide a unique tool to simultaneously probe the dust and gas content in the ISM of distant galaxies, independent of galaxies' intrinsic luminosities or star-formation rates. 
The study of QASs exploits sightlines to distant, luminous objects, such as quasars, that emit radiation over a wide-range of frequencies. As the quasar light passes through foreground galaxies, interstellar gas and dust 
produce absorption features in the quasar spectrum. The ISM of any galaxy with a luminous background source can be studied with this technique, but it is particularly effective when applied to the most
gas-rich systems, such as damped Lyman-$\alpha$ absorbers (DLAs; neutral hydrogen column densities N$_{HI}\geq2\times10^{20}$ cm$^{-2}$) and sub-DLAs (10$^{19}\leq$N$_{HI}<2\times10^{20}$ cm$^{-2}$). 
These gas-rich QASs are the primary neutral gas reservoir for star formation in the universe [e.g., \citet{Storrie00, Peroux05, Prochaska05}].
As opposed to flux-limited studies of gas and dust in emission, that are biased toward the most robustly star-forming objects, 
QASs provide insight into more quiescent or moderately-star-forming galaxies, potentially more similar to Local Group galaxies. 
 
Evidence for the presence of interstellar dust in gas-rich QASs has been found in both moderate-to-large surveys, as well as in investigations of individual systems. 
Studies of QASs show depletions of refractory elements and reddening of background quasars \citep{Pei91, Pettini97}, 
as well as a correlation between dust-reddening and gas metal absorption line strengths in absorbers at 1$<$z$<$2 \citep{York06}. 
The bulk of analyses exploring dust spectral properties in QASs, however, concentrate on signatures associated with the carbonaceous
dust component, such as the 2175~\AA~bump, rather than on the silicate dust, which exhibits prominent IR features near 10 and 18~$\micron$ [see, e.g., \citet{Draine03} for review of dust spectral signatures and their origins]. 
It is important to also consider this silicate dust, both because it is primarily produced by a different stellar sub-population, and because within the Milky Way
$\sim$66\% of the core mass of interstellar dust grains is silicate in composition \citep{Greenberg99, Zubko04}.

In an exploratory study, our group made the first measurement of silicate dust in a QAS. 
A $>$10-$\sigma$ detection of the 9.7~$\mu$m silicate absorption feature was made in the z$_{abs}=0.524$ DLA toward AO 0235+164 
using low-resolution \textit{Spitzer Space Telescope Infrared Spectrograph (IRS)} data. 
This was followed by identification of the 9.7~$\micron$~silicate absorption feature
in 5 additional strong QASs at redshifts 0.44$<z_{abs}<$1.31 \citep{Kulkarni11, Aller12}. 
Analyses of the silicate dust in these 6 QASs suggest both similarities and differences relative to that in the Milky Way ISM. 
In the Milky Way diffuse interstellar clouds, the peak optical depth of the silicate absorption is seen to scale with the extinction ($A_V$), and hence with the reddening 
[E(B-V)=$A_V/R_V$ where $R_V$, the selective extinction, is $\sim$3.1]. In the studied QASs, the silicate optical depth similarly scales with the reddening. However, 
the absorption features are 2-4 times deeper for a given reddening than those in the Milky Way \citep{Kulkarni11}. This disparity may stem from differences in dust composition, metallicity, and/or grain size distributions. 
A larger selective extinction may imply, for instance, larger dust grain sizes. Furthermore, in these 6 QASs comparisons of the peak wavelengths of the 10~$\micron$~silicate absorption feature, and of the substructure within
the feature, reveal variations suggestive of mineralogical and grain structural differences relative to one-another, and relative to the Milky Way ISM \citep{Kulkarni11,Aller12}.
These differences are most pronounced in the z=0.886 absorber toward PKS 1830-211 \citep{Aller12}, where the 10~$\micron$ absorption feature substructure is consistent with a crystalline olivine silicate dust grain origin, in 
marked contrast with the amorphous ($<$5\% crystalline) silicate dust grains in the Milky Way ISM \citep{Kemper04, Li07}. Exploring whether these trends and variations are found in other QASs is essential for quantifying
the evolution in dust grain properties. 

In this paper, we make progress toward this goal by presenting the first analysis of the 10 and 18~$\micron$ silicate absorption features in the z$_{abs}$=0.68466$\pm$0.00004 absorber \citep{Carilli93} 
toward the gravitationally lensed, z=0.944$\pm$0.002 blazar TXS 0218+357 at RA (J2000)= 02\fh21\fm05\fs5~and Dec (J2000) = +35\degr56\arcmin14\arcsec~ \citep{Patnaik92,Browne93,Cohen03}. 
Throughout our analysis we simply refer to this z$_{abs}$=0.6847 system, which is by far the strongest known metal absorption system along the quasar sightline, as \textit{the} TXS 0218+357 QAS. 
The TXS 0218+357 QAS was selected for this study for several reasons. 

First, this QAS is relatively rich in both gas and dust. The abundance of gas is evident from the large measured Mg~II absorption rest-frame equivalent width W$_{2796}=$3.0~\AA, the high
molecular hydrogen column density ($5\times10^{21} \leq N_{H2}\leq 5\times10^{23}$ cm$^{-2}$), and the detections of 21-cm absorption implying $N(H I) = 4\times10^{18} (T_s/f)$ cm$^{-2}$, 
where $T_s$ is the gas spin temperature and $f$ is the H~I covering factor \citep{Browne93,Kanekar02,Combes95,Menten96,Carilli93}. 
There are also signatures of dust in the QAS, including an exceedingly high measured reddening of the background quasar.
Statistical studies of mean DLA extinction curves using SDSS data find that most QASs at 1$<$z$<$5 have a
typical rest-frame reddening of E(B-V)$\sim$0.002 \citep{York06,Khare12}, while the TXS 0218+357 QAS exhibits a total measured E(B-V)=0.62$\pm$0.04, assuming R$_V=3.1$ \citep{Falco99}.   

Second, this QAS is known to be rich in molecules, similar to the PKS 1830-211 z=0.886 absorber. Since in the PKS 1830-211 QAS the interstellar silicate dust may be significantly crystalline \citep{Aller12}, unlike
that in the Milky Way, it is important to ascertain whether other molecule-rich QASs are likewise potentially rich in crystalline silicates. 
Molecules such as CO, CH, and H$_2$ trace cold, dense, star-forming material in the galaxy ISM.
The TXS 0218+357 QAS exhibits transitions for numerous molecules including 
CO, HCO$^+$, HCN, HNC, CN, CS, C$^{18}$O, $^{13}$CO, H$_2$O, NH$_3$, OH, H$_2$CO and its anti inversion, and the primordial LiH molecule which traces H$_2$ in low metallicity regions
 \citep{Wiklind95, Combes95, Combes97b, Combes98,Combes97,Gerin97,Menten96,Kanekar02,Kanekar03,Henkel05,Jethava07,Zeiger10,Friedel11}. 
Moreover, this absorber has among the highest molecular column densities at this redshift range \citep{Combes98}.
Studies of the QAS reveal that one of the gravitationally-lensed quasar sightlines passes through an optically thick and clumpy molecular cloud, while the other
sightline passes through a more optically thin region \citep{Combes95,Combes97b}. The derived column densities and abundances for molecules such as formaldehyde along the more obscured sightline are similar to those found in Milky Way molecular clouds \citep{Menten96}, while the NH$_3$ absorption measurements are more consistent with a diffuse cloud origin \citep{Henkel05}. 

Third, the z$_{abs}$=0.68466 QAS redshift  is optimal for studying both the 10 and 18~$\micron$ silicate absorption features with the \textit{IRS} instrument used in our previous studies. 
In the QAS rest-frame, the IRS spectra span the region 3.1-22.6~$\micron$, with adequate quasar continuum coverage between the absorption features. 
Additionally, for this quasar the \textit{IRS} sensitivity was sufficient to obtain data for this analysis in only 30 minutes of exposure time. 

Fourth, the z$_{em}$=0.944 \citep{Cohen03} background quasar is gravitationally lensed, which boosts its signal, and hence the signal-to-noise ratio (S/N) of the QAS gas and silicate dust absorption features. 
Furthermore, since this is a well-studied strong lens system, information about the QAS host galaxy topography, which is often lacking for other QASs, is available. 
Studies of the morphology of the lensing galaxy hosting the QAS indicate it is a gas-rich spiral galaxy with narrow absorption (Ca~II H\&K, Mg~II, Fe~II, Mg~I) and emission
([OII], H$\beta$, [OIII]) lines \citep{Carilli93,Browne93,Menten96,York05}, suggestive of some ongoing star-formation, comparable to that in spiral galaxies like the Milky Way based on the measured [OII] equivalent width \citep{Browne93}.
Radio studies [e.g., \citet{Odea92,Patnaik93}] reveal that the quasar is being lensed into two point sources and a small Einstein ring with an angular diameter of 335 mas 
($\sim$2.4~kpc in the lensing galaxy for H$_o$=70 km/s/Mpc). 
The optical and IR magnitudes along the two compact-object lines of sight imply a reddened and optically faint (V$>$21mag) source \citep{Lehar00}.

Lastly, owing to the well-studied strong gravitational lens, as well as to the monitored (variable) blazar, abundant ancillary data are available for this system. 
These data include spectral energy distributions (SEDs) extending from the $\gamma$ ray (e.g. Fermi, ROSAT, Swift) through the optical and NIR (e.g. HST), to the sub-mm (e.g. NRAO, PdBI) and radio (e.g., VLA, VLBA). 
However, while these data provide global constraints, owing to the blazar variability, predicting the flux at a given epoch is challenging. 

In this paper, we begin with a description of the \textit{IRS} observations, 
followed by a summary of the data reduction (\S\ref{IRSspect}), spectral extractions (\S\ref{Extract}), and quasar continuum normalizations (\S\ref{Norm}). We then
discuss the profile template fitting to the silicate absorption features produced by the TXS 0218+357 QAS (\S\ref{results}). 
We conclude with a discussion of our findings in comparison with the silicate dust in other QASs (\S\ref{disc}), followed by a summary of our main results (\S\ref{sum}). 
In the Appendices, we show the effects of different quasar continuum normalizations on our results (Appendix~\ref{normAP}), and discuss details of the template profiles applied to the data (Appendix~\ref{templateAP}). 
\section{DATA}\label{data}
Our data consist of spectra of the TXS 0218+357 quasar obtained using the \textit{IRS} instrument aboard the \textit{Spitzer Space Telescope}, as summarized in Table~\ref{ObsIT}. 
These data were observed as part of program 50783 (PI: V.P. Kulkarni) in the IRS staring mode. 
The blue peak-up filter with high peak-up accuracy was used to identify the quasar.
Five exposures were taken in each of the two nod positions at every available low-resolution spectral order, for a total of 10 exposures per order. 

\begin{deluxetable}{lccc}
\tabletypesize{\scriptsize}
\tablecaption{IRS Observations of TXS 0218+357}
\tablewidth{0pt}
\tablehead{
\colhead{Order} & \colhead{Obs. Frame} & \colhead{QAS Rest-Frame} & \colhead{$N_{exp} \times t_{exp}$} \\
\colhead{} & \colhead{$\micron$} & \colhead{$\micron$} & \colhead{sec}}
\startdata
SL2 & 5.2-7.7 & 3.1-4.6 &10x60s  \\
SL3 & 7.3-8.7 & 4.3-5.2 &10x60s  \\
SL1 & 7.4-14.5 & 4.4-8.6 & 10x60s  \\
LL2 & 14.0-21.3 & 8.3-12.6 & 10x30s  \\
LL3 & 19.4-21.7 & 11.5-12.9 & 10x30s  \\
LL1 &19.5-38.0 & 11.6-22.6 & 10x30s  \\
\enddata
\tablecomments{Details of the IRS spectra obtained for TXS 0218+357. In column 1 we list the spectral order following \textit{IRS} nomenclature from \citet{IRS}, followed by (column 2) the observed-frame spectral coverage; 
(column 3) the TXS 0218+357 QAS rest-frame coverage; and (column 4) the number of frames and exposure time per frame.}\label{ObsIT}
\end{deluxetable}

Although the quasar TXS 0218+357 is gravitationally lensed into a 335 mas diameter Einstein ring and two bright, compact components also separated by 335 mas \citep{Patnaik93},
our data do not resolve these individual sightlines through the foreground absorber. 
According to the IRS manual, the FWHM spatial profile widths obtained for calibration point sources (stars) exceed 2\farcs5 for every spectral order.
\textit{IRS} slits range in width from 3\farcs6 (SL2)-10\farcs7 (LL1) and in length from 57\arcsec -168\arcsec \citep{IRS}.  
Each \textit{IRS} spectrum, thus, is the superposition of the multiple sightlines through the QAS, and so we cannot investigate spatial variations in the galaxy ISM.

\subsection{Data Reduction}\label{IRSspect}

The 2-dimensional (2D) spectra obtained over the 40 total exposures were automatically reduced using the \textit{Spitzer IRS} Basic Calibrated Data (BCD) pipeline version S18.18.0.
This pipeline applied bias and dark subtractions, as well as flat-fielding the data. It also corrected for systemic effects such as droop and dark-current drift \citep{IRS}.

Prior to subtracting the background and extracting the quasar spectrum, we implemented two phases of bad-pixel cleaning (i.e. replacement). 
The background bad pixels were first cleaned using \textit{IRSCLEAN} (version 2.1), followed by the manual replacement of 
remaining anomalous background pixels and hot/cold pixels along the quasar spectrum using the \textit{Image Reduction and Analysis Facility (IRAF) epix} task.
Typically fewer than 6 pixels were replaced along the quasar spectrum in a single exposure.
Our tests find that our cleaning method results in a more conservative estimate of the significance of the structural parameters than would be obtained using the uncleaned, raw BCD data. 

The cleaned exposures in each nod position were combined using the \textit{Spitzer} \textit{IDL} script \textit{coad}.
Each pixel value in the combined frame was computed as the trimmed mean of the pixels in the input images. Pixels with no value
and those fatally flagged in the associated mask files were excluded. The corresponding pixel uncertainty was computed as the square root of the sum of the squares in the input uncertainty
file, normalized by the number of pixels used to calculate the trimmed mean. 

We subtracted the background from each of the coadded frames using the nod exposures.  
A background-subtracted 2D spectrum was produced by subtracting the nod frames from one another. The associated uncertainty files were propagated in quadrature. 

\subsection{Spectral Extraction}\label{Extract}

The wavelength-calibrated, 1D quasar spectrum was extracted from the background-subtracted, 2-D spectrum for each nod using the \textit{SPICE GUI} (version 2.5). Each extraction was run using the generic point-source extraction mode
with the optimal extraction setting. The optimal extraction weights each pixel in the aperture using the S/N of the pixel, as computed using the associated pipeline uncertainty file. This method improves the S/N for 
intrinsically low-S/N spectra.

The spectra from the two nods were combined by averaging the flux values at every wavelength. The associated uncertainties are a combination in quadrature of the formal uncertainty associated with the mean 
(determined by propagating the pipeline uncertainties) and the scatter of the flux measurements about the mean at a given wavelength. 
In general, data points in the final spectrum with large error bars stem from discrepancies between the two nod spectra. 
These differences may originate from a slightly hot pixel along the quasar spectrum or from an emission feature which lies on a cold pixel in one nod. Rather than eliminating these data points or simplistically
assuming them to originate in hot pixels, we conservatively retain them. 
These noisy points have a minimal impact on our absorption feature fitting, because their large uncertainties effectively de-weight the data points.

The nod-averaged spectra (SL2, SL1, LL2, LL1) were stitched together to produce a single contiguous spectrum. A few ($\leq$3) anomalous data points at each  
spectral segment edge were removed to produce a smoother joining. 
All data at $\lambda\geq$35.8~$\micron$ were rejected, as they showed significant degradation. The SL2 and SL1 data were scaled by multiplicative factors to match the identically
cleaned and extracted overlap (SL3) data.
A similar process was repeated for scaling the LL1 and LL2 orders to match the the LL overlap (LL3) data, and scaling the LL orders to match the SL orders. 
The final multiplicative rescaling factors for the SL2, SL1, LL2, and LL1 orders were 0.82, 0.92, 1.14, and 1.05, respectively.
Since the resultant quasar spectrum is continuum normalized prior to our analysis of the silicate absorption features, as described below, the absolute flux normalization is not crucial. 

\begin{figure}
\epsscale{1.}
\plottwo{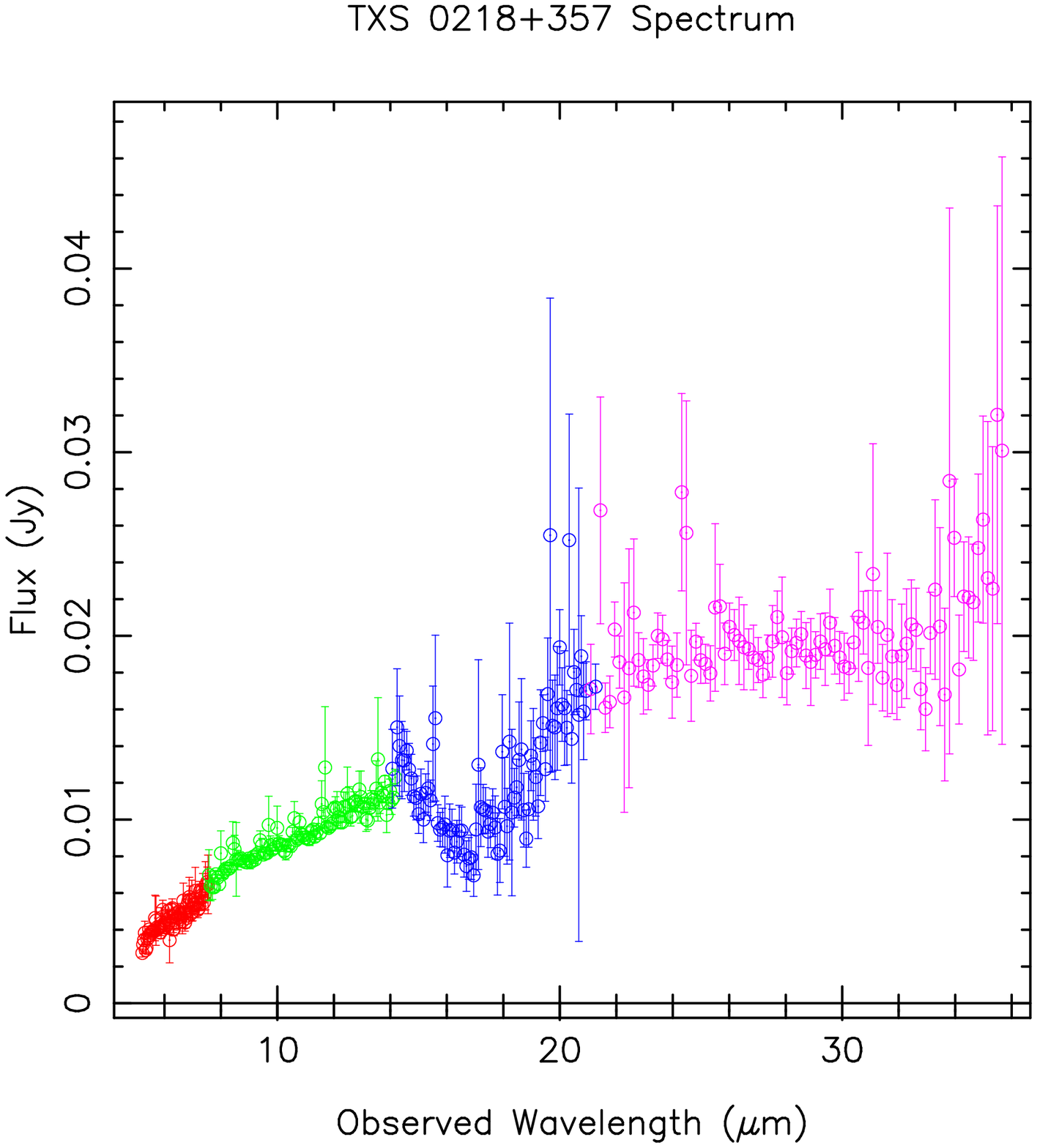}{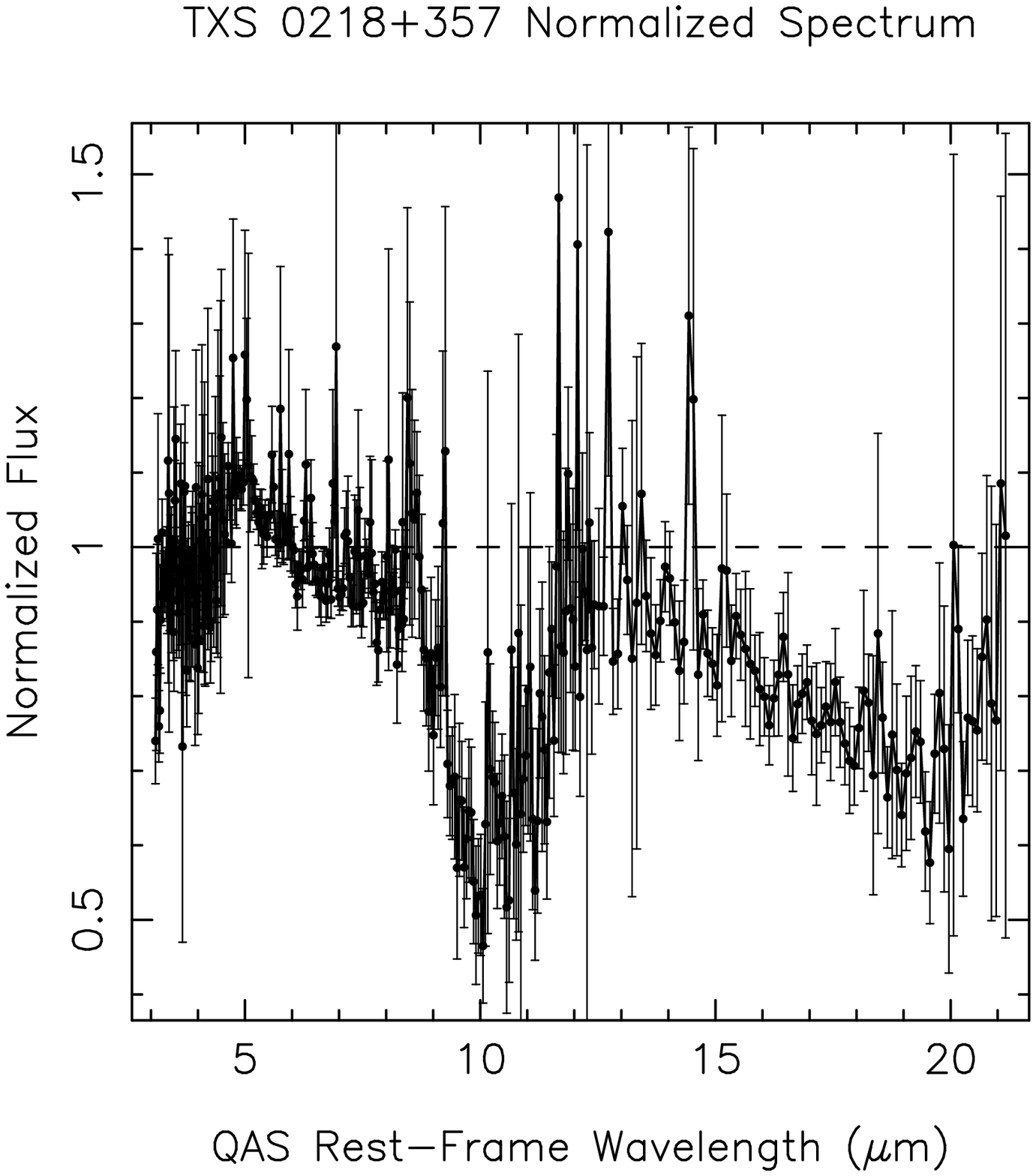}%
\caption{\textit{Left:} TXS 0218+357 flux density as a function of the observed wavelength. The colors denote different spectral orders as follows: red-SL2, green-SL1, blue-LL2, and magenta-LL1.
\textit{Right:} The quasar-continuum-normalized TXS 0218+357 spectrum as a function of the QAS rest-frame wavelength. The dashed horizontal line denotes the normalized continuum. 
The relatively strong 10~$\micron$ silicate absorption feature is apparent, as is a broader and shallower 18~$\micron$ silicate absorption feature.} \label{spect}
\end{figure}

The overlapping data points at the order joinings were averaged to produce a single flux at each wavelength. The errors are a combination of the propagated measurement uncertainties and the standard deviation about the mean for each data point.
A visual inspection indicated that the uncertainties in the SL1/SL2 and LL1/LL2 overlap regions were typically larger than those in the associated overlap spectra (SL3, LL3). Therefore, in regions where an overlap spectrum flux value was
available, it was used in place of the first and second order data. Tests indicate that these replacements slightly reduce the mean uncertainties in the final spectrum, while not impacting the shape of the absorption 
features or quasar continuum. The final combined spectrum is shown in Figure~\ref{spect} (left), as a function of the observed wavelength. We do not bin the data in this analysis, as was done in the analyses of several previous 
QASs \citep{Kulkarni07, Kulkarni11}, because while binning reduces the data point uncertainties, it may also obscure substructure in the absorption features. 

The combined, unbinned quasar spectrum was finally shifted to the z=0.68466 redshift QAS rest-frame. This QAS rest-frame spectrum was used in all subsequent steps of the analysis, including fitting the quasar continuum and 
performing the mineralogical fits to the silicate absorption features. All wavelengths referred to in the following sections are in the QAS rest-frame, unless explicitly stated otherwise.

\subsection{Quasar Continuum Normalization}\label{Norm}

We normalized the combined spectrum by the quasar continuum, prior to measuring the silicate absorption features. The physical shape of the quasar continuum is unknown. 
Therefore, we fitted a series of low-order functions to our data, including linear, power-law, 3rd order Chebyshev polynomial, and cubic spline fits. We experimented with several different fitting regions
and data point exclusions for each fit, as detailed in Appendix~\ref{normAP}. We selected as the best-fitting a 3rd order Chebyshev polynomial fitted by excluding the 8.5-12.5~$\micron$ and 16.5-20.4~$\micron$ regions
containing the silicate absorption features, and manually rejecting several data points with anomalously large uncertainties and deviant flux values in the LL1/LL2 joining region. 
This fit visually matches the data slightly better than the other considered quasar continuum fits, although \textit{all} normalizations result in similar 10~$\micron$ silicate feature fits,
as detailed in Appendix~\ref{normAP}. The largest discrepancies occur for the 18~$\micron$ absorption feature, which is near the `red' end of our covered spectral range. Depending
on the shape of the adopted fit, the 18~$\micron$ feature may be relatively broad and deep, or may be narrow and shallow. Since there are no mid-IR (\textit{Spitzer IRAC, MIPS}; \textit{Herschel PACS})
archival data to constrain the slope of the continuum beyond our spectrum, and far-IR/radio data obtained in different epochs cannot be compared reliably (due to the temporal variability of the blazar),
we cannot determine the physical quasar continuum shape near 18~$\micron$. Therefore, we do not emphasize our analysis of the 18~$\micron$ absorption feature.
The final quasar-continuum-normalized spectrum we obtained is depicted in Figure~\ref{spect} (right). 
\section{SILICATE ABSORPTION RESULTS}\label{results}
An examination of the continuum normalized spectrum (Figure~\ref{spect}, right) reveals a detection of both relatively strong 10~$\micron$ silicate absorption, 
as well as a broader and shallower 18~$\micron$ silicate feature. We measured the rest-frame equivalent width (EW) of the 10~$\micron$ feature to be 0.83$\pm$0.08~$\micron$,
which corresponds to a 10.7$\sigma$ detection significance. 
We conservatively estimate the equivalent width uncertainty to be the combination (in quadrature) of the photon noise and the continuum noise, with the continuum noise dominating.
To estimate the continuum uncertainty, we used the prescription of \citet{Sembach92}, with a ``nudge factor" of 0.3. 
This equivalent width is nearly quadruple that of the 10~$\micron$ absorption feature in the z=0.524 QAS toward AO 0235+164, and is almost double that of the largest 10~$\micron$ equivalent width
QASs previously studied \citep{Kulkarni11}. 

We similarly measured the EW of the 18~$\micron$ silicate feature to be 1.40$\pm$0.35~$\micron$; a 4.1$\sigma$ significance detection.
We caution, however, that the continuum normalization near 18~$\micron$ remains uncertain because of the absence of longer-wavelength data. This feature may be over- or under-estimated. 
If we assume that the absorption profile is produced by amorphous olivine, as we discuss below, then this equivalent width is a lower limit, because the 
predicted 18~$\micron$ feature extends past the `red' edge of our data. 

\subsection{Template Fitting Methodology}
In order to determine the peak optical depth and to place some constraints on the mineralogical composition and grain properties of the dust producing the silicate absorption features, we fitted template profiles
to the normalized quasar spectrum. We follow the methodology outlined in \citet{Aller12}. 
The transmitted flux at any frequency $\nu$ is given by f$_{\nu}$/f$_o=\exp[-\tau_{\nu}]$, where $\tau_{\nu}$ is the measured optical depth at frequency $\nu$, assuming a constant covering factor ($C_f$) of 1.0. 
The optical depth can be expressed in terms of the peak optical depth ($\tau$) and the template unity-normalized optical depth profile ($f_{\nu t}$) as $\tau_{\nu}\equiv\tau\cdot f_{\nu t}$.
The optimal peak optical depth is determined by minimizing the chi-squared over the specified fitting region. 
The 1-$\sigma$ uncertainty is determined from an interpolation of the high and low $\tau$ values,where $\Delta\chi^2$=1.0 relative to the minimum measured
chi-squared. We determine which mineral best fits the observed silicate absorption profile by comparing the reduced chi-squared ($\chi_r^2$) values
and the chi-squared probabilities, which quantify the probability that a similar or larger $\chi^2$ would be obtained by chance. 

We considered a representative sample of silicate template profiles obtained from the literature for laboratory minerals and for astrophysical samples of SiC. 
These laboratory templates include both amorphous olivines (Mg$_{2x}$Fe$_{2-2x}$SiO$_4$) and pyroxenes (Mg$_x$Fe$_{1-x}$SiO$_3$) and crystalline olivines and pyroxenes, 
with a range of chemical compositions and grain temperatures, shapes, porosities, and sizes. 
We explore multiple templates in each mineral category because studies [e.g., \citet{Chiar06,Speck11}] have established that physical variations in these grain properties, 
as well as the laboratory methodology employed to measure the mineral properties for a given species, can significantly impact
the silicate profile shape and peak absorption wavelengths. In our analysis, however, we only describe the results for the best-fitting (lowest $\chi_r^2$ and highest probability)
profile fit in each mineral category. All of the template profiles have been measured for, or utilized in, comparison with Galactic astrophysical dust sources, and are discussed in more detail in Appendix~\ref{templateAP},
and in \citet{Aller12}.

We emphasize that our goal in exploring these template profile fits is \textit{not} to unambiguously determine the mineral 
and dust grain properties which explain the observed silicate dust absorption. This would require much higher S/N data. 
Rather, we use the templates to place some constraints on the dust grain properties, by exploring whether certain classes of minerals are viable 
and whether certain classes of minerals can be ruled-out as plausible origins for the observed silicate features. We explore several minerals in each category
solely to ensure that an entire mineral category is not ruled out simply because one or two species of minerals in the category fit the data poorly.
 
In addition to considering the single mineral fits, we also briefly consider the possibility that the silicate dust may be a mixture of two different minerals or grain structures. \citet{Spoon} found that silicate dust in ULIRGs was best explained
by a mixture of amorphous and crystalline olivines. Likewise, there is no reason why the interstellar dust in the TXS 0218+357 QAS must be homogeneous. We implement these two-mineral fits by determining 
$\tau_{\nu,1}$ and $\tau_{\nu,2}$ in the expression f$_\nu$/f$_o=\exp[-(\tau_{\nu,1}+\tau_{\nu,2})]$ again assuming a constant covering factor of $C_f=1.0$, following the methodology described in \citet{Aller12} 
\subsection{10~$\micron$~ Fits}
\begin{figure}
\epsscale{1.}
\plotone{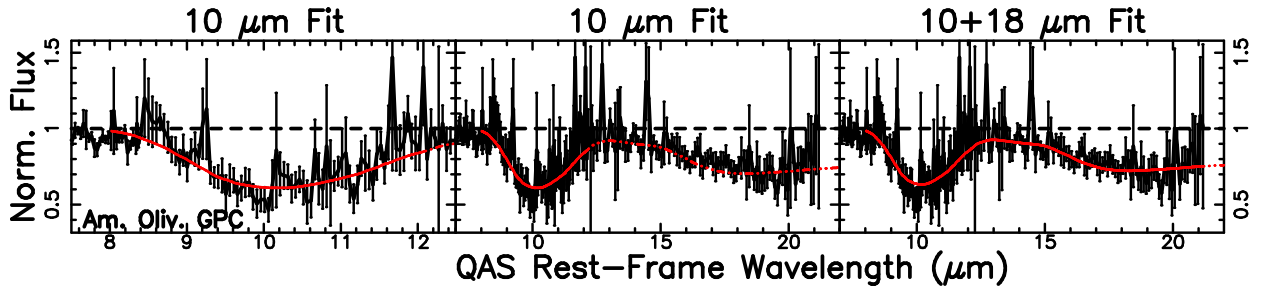}%
\caption{Comparison of quasar-continuum normalized spectrum with best-fitting, amorphous olivine template profile (in red). \textit{Left and Center:} Profile fit over 10~$\micron$ feature only; \textit{Right:} Profile fit over 10 and 18 $\micron$ regions
simultaneously. The profile fit is shown as a solid line over the regions where fitting (i.e. $\chi^2$-minimization) occurred, and as a dash-dot-dot-dotted line over all other wavelengths. The shape, breadth, and peak absorption wavelengths
of the laboratory amorphous olivine template 10 and 18~$\micron$ silicate features provide a good match to the data. [See the electronic edition of the Journal for a color version of this figure.]} \label{BESTSINGLEFIT}
\end{figure}

\begin{figure}
\epsscale{1.}
\plotone{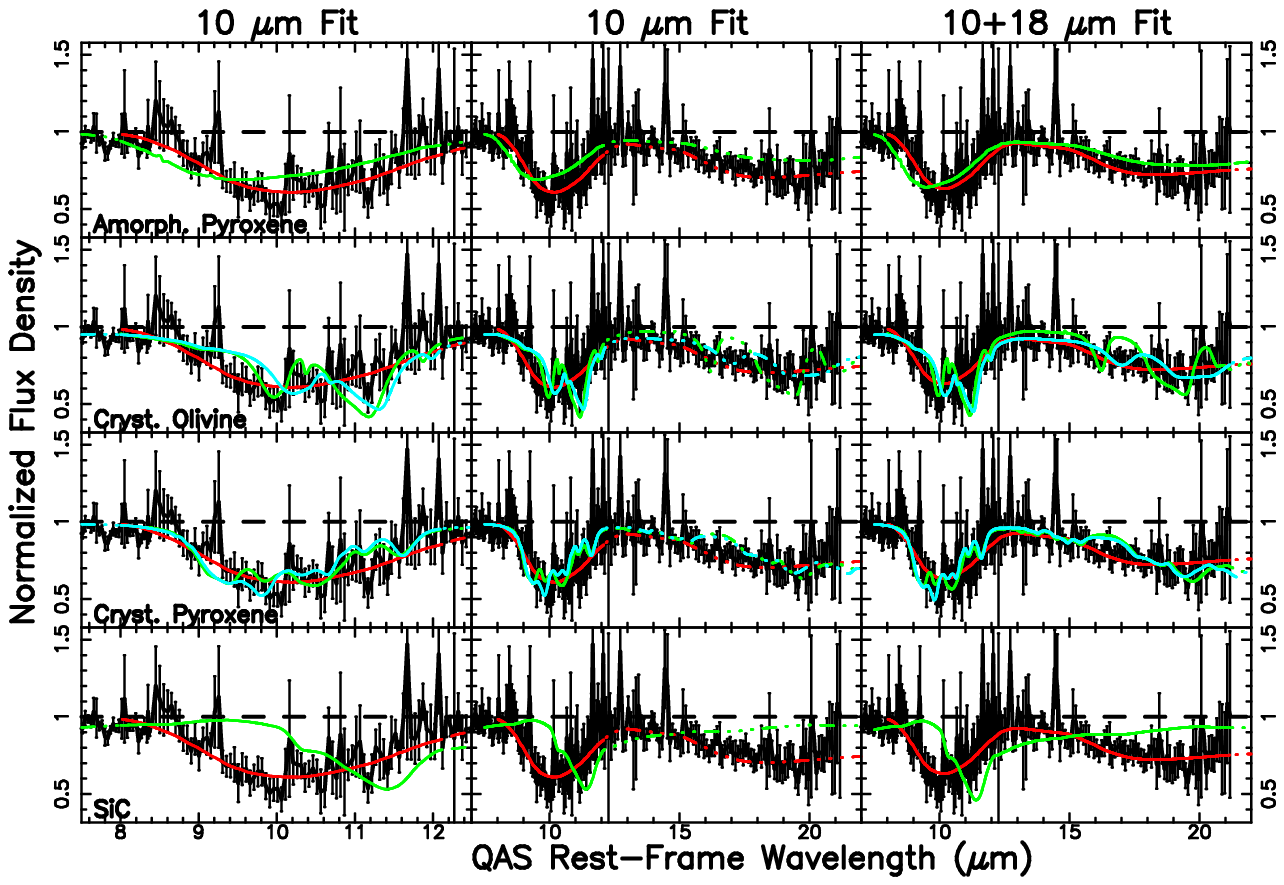}%
\caption{Similar to Figure~\ref{BESTSINGLEFIT}, but depicting the minerals which produce the best fit in each of the mineral categories summarized in Tables~\ref{FITS-10}-\ref{FITS-10+18}.
The mineral species producing the best-fit over the 10~$\micron$-only fitting region are shown in green, and those which provide the best-fit over the expanded 10 and 18~$\micron$ fitting
regions are shown in cyan, if they differ. The overall best-fitting amorphous olivine template is shown in red in every panel, for comparison. We note that SiC provides a poor fit over both the 10 and 18~$\micron$ features.} \label{SINGLEFIT}
\end{figure}

We commence with fitting the template profiles to \textit{only} the TXS 0218+357 QAS 10~$\micron$ silicate absorption feature. We restrict fitting (i.e. $\chi^2$-minimization) to the 
$8.0\leq\lambda\leq12.1~\micron$ region in which the absorption occurs. This relatively narrow fitting region is adopted to prevent undue weighting from other, weaker spectral features which 
may exist within the quasar continuum. Small changes in the breadth of the fitting region have little impact on our results. The fitting results are summarized in Table~\ref{FITS-10} and
the fits are illustrated in the left panels of Figures~\ref{BESTSINGLEFIT}-\ref{SINGLEFIT}. 

The best fit is produced by a species of amorphous olivine with porous, ellipsoidal dust grains (Amorph.Oliv.GPC), with a measured peak optical depth for the 10~$\micron$ feature of $\tau_{10}$=0.49$\pm$0.02. As illustrated 
in the left panel of Figure~\ref{BESTSINGLEFIT}, this profile well matches the depth, breadth, and peak location of the silicate feature. The best mineral fits in the other mineral categories are inferior
both statistically (Table~\ref{FITS-10}) and visually (Figure~\ref{SINGLEFIT}). 

Amorphous pyroxene and SiC have both been exploited to explain silicate dust absorption in Milky Way environments such as near AGB stars [e.g. \citet{Speck09,Speck11,Messenger13}]. 
However, we find, for our data on the TXS 0218+357 QAS, that the best-fitting amorphous pyroxene (Amorph.Pryox.GPC; porous, ellipsoidal grains) and SiC (green $\alpha$-SiC) templates explored produce particularly poor fits to the data
($\chi^2$ fit probabilities consistent with 0\%). The fits neither match the peak wavelength nor the breadth of the observed absorption feature. 

Crystalline olivines and pyroxenes have been utilized to explain silicate absorption features in both Galactic [e.g., \citet{Molster05}]  and extragalactic [e.g., \citet{Spoon,Aller12}] dust.
Our best-fitting crystalline olivine (T=10K low-temperature forsterite; Mg$_{2}$SiO$_4$) and crystalline pyroxene (orthobronzite; Mg$_{0.88}$Fe$_{0.12}$SiO$_3$) fits are mildly suggestive that some crystalline material could be present, 
but the fits are poorer than for amorphous olivine (see Table~\ref{FITS-10}). Although some features expected for these minerals align
with substructure in our absorption profile, such as the small central ridge in the 10~$\micron$ absorption feature, all of the substructure is our data is well-within the noise uncertainty limits, 
and so does not provide compelling evidence for crystallinity. 

If we instead used the Galactic astrophysical profile templates discussed in \citet{Kulkarni11} to fit the data, our peak optical depth for the 10~$\micron$ feature would be largely unchanged. Of the three templates, 
the diffuse cloud $\mu$Cephei red supergiant \citep{Roche84} template produces the best match to our data, with $\tau_{10}=0.50^{+0.03}_{-0.02}$ and a $\chi_r^2=0.96$ (P=59\%). 
While the optical depth is consistent with that derived using the laboratory amorphous olivine template, the fit quality is poorer. The GCS-3 Galactic center diffuse ISM source \citep{Spoon,Chiar06}
also provides an adequate fit ($\chi_r^2$=1.09, P=26\%) with a consistent optical depth ($\tau_{10}=0.51\pm0.03$). The only template discussed in \citet{Kulkarni11} which does
not provide a good fit to our data is the Trapezium molecular cloud \citep{Forrest75,Bowey01} template which has a $\chi_r^2>2$ fit ($\chi^2$ fit probability consistent with 0\%). Based on these astrophysical templates,
we would conclude that the TXS 0218+357 QAS is not well matched by Galactic molecular clouds, but is broadly consistent with material in other, more diffuse, Galactic environments.

\begin{deluxetable}{lllllll}
\tabletypesize{\scriptsize}
\tablecaption{Template Profile Fits to TXS 0218+357 QAS 10~$\micron$ Feature}
\tablewidth{0pt}
\tablehead{
\colhead{Category} & \colhead{Mineral} & \colhead{$\tau_{10}$} & \colhead{$\chi_r^2$} & \colhead{P} & \colhead{n$_{pts}$} & \colhead{$\lambda_{min}-\lambda_{max}$}}
\startdata
Amorph. Olivine (Best) & Amorph.Oliv.GPC (1)& 0.49$\pm$0.02 & 0.86 & 81.8 & 85 & 8.02-12.10 \\
Amorph. Pyroxene & Amorph.Pyrox.GPC (1) & 0.37$\pm$0.02 & 1.77 & 0.0 & 86 & 8.00-12.10 \\
Cryst. Olivine & Forsterite(T=10K)(2) & 0.88$\pm$0.05 & 1.49 & 0.2 & 86 & 8.00-12.10 \\
Cryst. Pyroxene & Bronzite (3) & 0.53$\pm$0.03 & 1.08 & 28.7 & 77 & 8.34-12.10\\
SiC & G0-Green-aSiC (4) & 0.63$\pm$0.05 & 4.99 & 0.0 & 86 & 8.00-12.10\\
\enddata
\tablecomments{For each mineral category, we present the mineral species which produced the best fit. 
We list the (column 1) mineral category; (column 2) mineral (and reference); (column 3) 
measured peak optical depth of the $\sim$10~$\micron$ feature assuming $C_f=1.0$; (column 4) reduced chi-squared; (column 5) percentage chi-squared probability; (column 6) 
the number of points used for the fit; and (column 7) wavelength range over which the fit was performed (in $\micron$). We use the term 
$\sim$10~$\micron$ peak feature to denote the location of greatest absorption in the template profile; for some minerals this is not 
precisely at 9.7 or 10~$\micron$, as detailed in Table 10 of \citet{Aller12}.
References: (1)  \citet{Chiar06} based on data from \citet{Henning99}; (2) \citet{Koike06}; (3) \citet{Jaeger} with tabulated data provided by J\"{a}ger; \& (4) \citet{Friedemann}.} \label{FITS-10}
\end{deluxetable}

\subsection{10~$\micron$ \& 18~$\micron$~ Fits}
Given the uncertainty associated with the quasar continuum normalization near 18~$\micron$, we do not explicitly fit template profiles to this feature alone. We instead 
consider the 18~$\micron$~silicate absorption feature contribution in two ways. First, we examine the predicted shape of the 18~$\micron$~feature based on the derived 10~$\micron$~peak optical 
depth. Second, we consider fits in which the $\chi^2$-minimization is implemented over $8.0\leq\lambda <22.0~\micron$,
thus simultaneously including the 10~$\micron$ and the 18~$\micron$ features in determining the best mineral fits and peak optical depths. 

We illustrate the predicted 18~$\micron$~feature, based on our previously described 10~$\micron$ fits, in the central panels of Figures~\ref{BESTSINGLEFIT}-\ref{SINGLEFIT}.  
The predicted amorphous olivine 18~$\micron$~absorption feature is a good match
to the observed spectrum. The predicted turnover between 10 and 18~$\micron$ follows the data, and both the peak and breadth of the feature are well-aligned with the prediction. This both provides additional
support for the presence of amorphous olivine at a measured $\tau_{10}=0.49\pm0.02$ peak optical depth, and lends credence, a posteriori, to our adopted quasar continuum normalization. 
Once again, SiC and amorphous pyroxene provide overall poor predictions of the shapes and locations of observed 10 and 18~$\micron$ features. 
The crystalline olivine and crystalline pyroxene fits match some of the low-level substructure near 18~$\micron$,
although this substructure is well within the noise limits, but other predicted features such as a rise near 16-17~$\micron$ are absent or muted in the data. 

We next extended the fitting to cover both the 10 and 18~$\micron$ regions simultaneously, the results of which are detailed in Table~\ref{FITS-10+18}. 
These fits are depicted in the rightmost panels of Figures~\ref{BESTSINGLEFIT}-\ref{SINGLEFIT}.
We find that over this expanded fitting region, the best-fitting mineral remains the amorphous olivine species with porous, ellipsoidal grains. The derived peak optical depth of this fit ($\tau_{10}=0.46\pm0.02$) is consistent
with that derived when fitting over only the 10~$\micron$ region. Amorphous pyroxene and SiC still produce poor fits. 
The largest differences occur for the crystalline olivines and pyroxenes. Different species are identified when the 18~$\micron$~silicate
absorption feature is included in the fit: hortonolite (Mg$_{1.1}$Fe$_{0.9}$SiO$_4$) and natural orthoenstatite (Mg$_{0.96}$Fe$_{0.04}$SiO$_3$), respectively.
In Figure~\ref{SINGLEFIT} we illustrate the species of crystalline olivine and pyroxene which
produce superior fits over this extended fitting region in cyan. It is apparent that while these species provide slightly poorer fits over the 10~$\micron$ feature than those originally identified, they are also slightly
less structure-rich over the 18~$\micron$ feature, which is more consistent with the observed data. Higher S/N and higher resolution data would be required to confirm these features
and identify the most viable crystalline species. 

\begin{deluxetable}{lllllll}
\tabletypesize{\scriptsize}
\tablecaption{Template Profile Fits to TXS 0218+357 QAS 10 \& 18~$\micron$ Features}
\tablewidth{0pt}
\tablehead{
\colhead{Category} & \colhead{Mineral} & \colhead{$\tau_{10}$} & \colhead{$\chi_r^2$} & \colhead{P} & \colhead{n$_{pts}$} & \colhead{$\lambda_{min}-\lambda_{max}$}}
\startdata
Amorph. Olivine (Best) & Amorph.Oliv.GPC (1)& 0.46$\pm$0.02 & 0.70 & 99.9 & 179 & 8.02-21.17 \\
Amorph. Pyroxene & Amorph.Pyrox.GPC (1) & 0.44$\pm$0.02 & 1.34 & 0.1 & 180 & 8.00-21.17 \\
Cryst. Olivine & Hortonolite\tablenotemark{a} (2)& 0.80$\pm$0.03 & 1.18	 & 5.2 & 180 & 8.00-21.17\\
Cryst. Pyroxene & Nat.Enstatite\tablenotemark{a} (2) & 0.71$\pm$0.03 & 1.00 & 47.4 & 180 & 8.00-21.17\\
SiC & G0-Green-aSiC (3) & 0.77$\pm$0.04 & 3.96 & 0.0 & 180 & 8.00-21.17\\
\enddata
\tablenotetext{a}{Different species of mineral was best in mineral category in Table~\ref{FITS-10}.}
\tablecomments{Similar to Table~\ref{FITS-10}, but listing the peak optical depths derived
when the fitting-region, i.e. the wavelength range over which the $\chi^2$-minimization was performed, is expanded to include the 18~$\micron$ region.
References: (1)  \citet{Chiar06} based on data from \citet{Henning99}; (2) \citet{Jaeger} with tabulated data provided by J\"{a}ger; \& (3) \citet{Friedemann}.} \label{FITS-10+18}
\end{deluxetable}

\subsection{Two Mineral Fits to 10~$\micron$~Feature}\label{twofitsect}
\begin{figure}
\epsscale{0.9}
\plotone{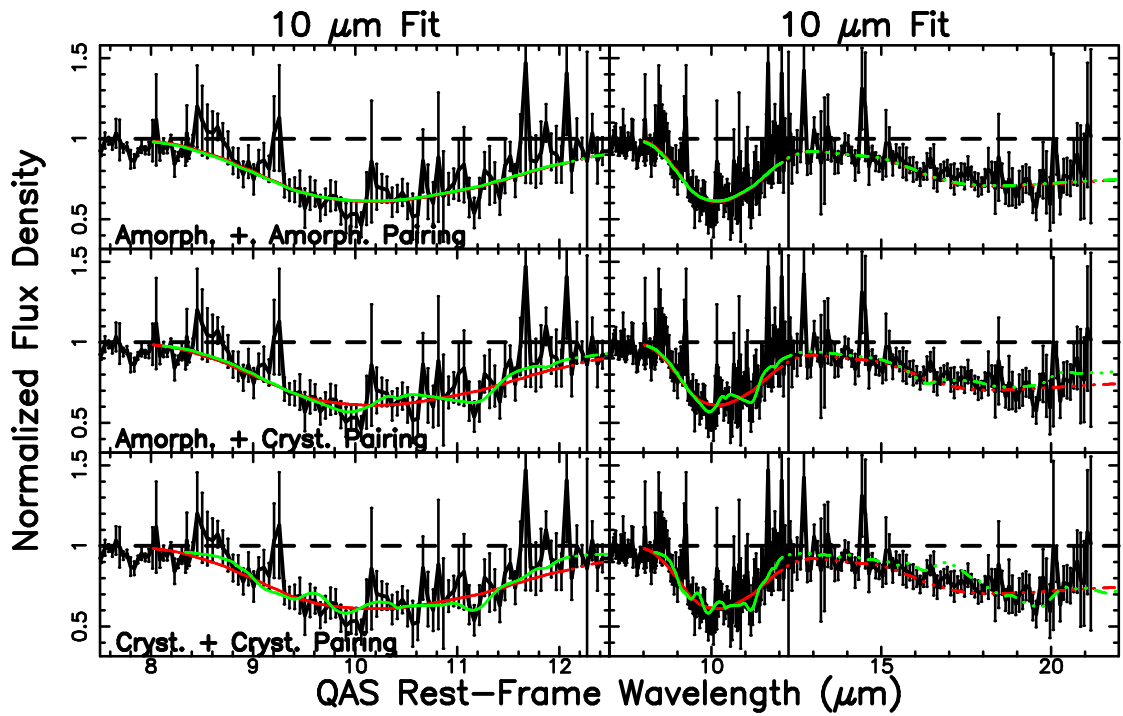}%
\caption{Comparison of quasar-continuum normalized spectrum with best-fitting, single amorphous olivine template profile (in red) and best-fitting mineral pairings (in green), 
each fit over the 10~$\micron$ silicate feature. The fit-prediction over the 18~$\micron$ region is shown in the right panel. A solid line is used for the profile 
over the regions where fitting occurred, and as a dash-dot-dot-dotted line over all other wavelengths.} \label{TWOFIT}
\end{figure}

Lastly, we briefly investigate the possibility that a blend of minerals produces the observed 10~$\micron$~silicate absorption feature. For simplicity, we consider pairings of all of the minerals identified 
in Tables~\ref{FITS-10}-\ref{FITS-10+18}, as well as pairings containing several additional amorphous and crystalline olivine species. 
The results of the two-mineral fitting, performed over only the 10~$\micron$~feature, are summarized in Table~\ref{FITS-TWO} and illustrated in green in Figure~\ref{TWOFIT} . 
In every panel we include the best-fitting amorphous olivine fit (in red) for visual reference. 

We find that the best overall fit is produced by a combination of crystalline olivine (San Carlos olivine, T=100K) and crystalline pyroxene (orothobronzite) with measured peak optical depths of 
$\tau_{oliv}$=0.36$\pm$0.07 and $\tau_{pyrox}$=0.34$\pm$0.04, respectively. We obtain a $\chi_r^2=0.71$, which is slightly lower than that obtained using the single mineral amorphous olivine (GPC) template fit ($\chi_r^2=0.86$). 
This results in a slightly higher (97\% versus 82\%) probability for the fit. However, as illustrated in Figure~\ref{TWOFIT}, although the crystalline olivine-crystalline pyroxene blend has a relatively high  
fit probability, it also predicts many small oscillations, or substructure features, within the 10 and 18~$\micron$ absorption features. This substructure
cannot be ruled out, but we favor the smoother prediction of amorphous olivine, due to the lack of compelling evidence for substructure in the observed profile at the S/N of our data.
The addition of crystalline olivine to an amorphous olivine with solid, spherical particles (Amorph.Oliv.GS) also formally improves the quality of the fit over a pure amorphous olivine fit (95\% versus 82\%).
Although, again, without higher S/N data it is impossible to determine whether the associated small substructures being fit have a physical origin or whether they stem from noise in the data. 
The combination of amorphous pyroxene with amorphous olivine does not lead to an improvement in the fit.

\begin{deluxetable}{lllllllll}
\tabletypesize{\scriptsize}
\tablecaption{Two Mineral Template Profile Fits to TXS 0218+357 QAS 10~$\micron$ Feature}
\tablewidth{0pt}
\tablehead{
\colhead{Category} & \colhead{Mineral 1} & \colhead{Mineral 2} & \colhead{$\tau_{1}$} &  \colhead{$\tau_{2}$} & \colhead{$\chi_r^2$} & \colhead{P} & \colhead{n$_{pts}$} & \colhead{$\lambda_{min}-\lambda_{max}$}}
\startdata
Amorph.+Amorph. & Amorph.Pyrox.GPC (1) & Amorph.Oliv.GPC (1) & 0.02$^{+0.04}_{-0.02}$ & 0.47$^{+0.04}_{-0.05}$ & 0.87 & 80.4 & 85 & 8.02-12.10\\
Amorph.+Cryst. & Amoph.Oliv.GS (1) & SanCarl.Oliv. (2) & 0.38$\pm$0.05 & 0.28$\pm$0.08 & 0.78&93.2&83&8.12-12.10\\
Cryst.+Cryst. (Best) & Bronzite (3) & SanCarl.Oliv. (2) &0.34$\pm$0.04 &0.36$\pm$0.07 &  0.71 & 97.2 & 77 & 8.34-12.10\\
\enddata
\tablecomments{We present the best amorphous+amorphous, amorphous+crystalline, and crystalline+crystalline mineral pairings using the minerals identified in Tables~\ref{FITS-10}-\ref{FITS-10+18}, supplemented
by several additional species of amorphous and crystalline olivines.
For each fit we list the (column 1) mineral category; (columns 2-3) minerals (and references); (columns 4-5) peak optical depths at the $\sim$10~$\micron$ feature assuming $C_f=1.0$; (column 6) 
reduced chi-squared; (column 7) percentage chi-squared probability; (column 8) number of points used for the fit; and (column 9) wavelength range over which the fit was performed (in $\micron$).
These mineral pairing fits are shown in green in Figure~\ref{TWOFIT}.
References: (1)  \citet{Chiar06} based on data from \citet{Henning99}; (2) \citet{Koike06}; \& (3) \citet{Jaeger} with tabulated data provided by J\"{a}ger.}\label{FITS-TWO}
\end{deluxetable}

\section{DISCUSSION AND CONCLUSIONS}\label{disc}
We find that the best-single mineral fit to the silicate absorption features in the TXS 0218+357 QAS is produced by a laboratory amorphous olivine silicate template with 
a  measured peak optical depth of $\tau_{10}=0.49\pm0.02$. This relatively high optical depth is not a product of our adopted best-fitting template shape. 
All of the mineral fits over both the 10~$\micron$ feature and combined 10 and 18~$\micron$ features
with $\chi_r^2\sim$1 result in a measured $\tau_{10} \gtrsim 0.45$. For the two mineral pairings, the combined 10~$\micron$ peak optical depths are also consistently high.
This derived peak optical depth for the TXS 0218+357 QAS is significantly larger than those measured for any of the 6 previously studied QASs, as illustrated in Table~\ref{COMPARE}.
Despite the deeper absorption, however, the best-fitting mineral species is consistent with several other QASs and with silicate dust grains in the Milky Way ISM. 
We discuss the implications of this high level of absorption and the dust grain species in the following sections. 

In comparing the peak optical depth measurements from the different QASs we now use both the measured peak optical depth values ($\tau_{10,obs}$), 
as well as the optical depths corrected ($\tau_{10,cor}$) by accounting for the covering factor ($C_f$). 
This correction has been implemented using the expression 
\begin{equation} 
\tau_{cor}=-\ln \left( 1 - {1-e^{-\tau_{obs}} \over C_f} \right) 
\end{equation} 
The covering factor quantifies the fraction of the subtended area along the sightline to the continuum source obscured by the dusty material, and in many systems may be less than 1.0, thus implying a higher physical optical depth 
than that measured. Since direct estimates of the dust covering factor along the sightline are not feasible, we assume that the dust traces either the molecular or H~I gas, and has a similar covering factor. For the TXS 0218+357 QAS, 
both molecular line analyses [e.g., \citet{Muller07}] and H~I studies 
[e.g., \citet{Patnaik95}] find a difference between the measured fluxes for the southwest and northeast lensed quasar images, with a flux ratio of $4.2^{+1.8}_{-1.0}$ at 106 GHz \citep{Muller07}. 
However, the precise value of the gas covering factor is debated in the literature [e.g., \citet{Jethava07,Zeiger10}], and the value at 10~$\micron$ may differ from that at $\sim$100~GHz. It has also been suggested that 
the H~I absorption regions may be more extended than the region producing the molecular lines \citep{Menten96}. For this analysis we adopt a value 
for the covering factor of $C_f=0.8$, which is broadly consistent with values found by \citet{Wiklind95} for molecular material in this system. This results in a corrected optical depth value
of $\tau_{10,cor}=0.67\pm0.04$ for the TXS 0218+357 QAS. For the z=0.886 QAS toward the lensed quasar PKS 1830-211, we adopt a lower total covering factor of $C_f=0.35$ \citep{Frye97,Henkel08}, which
raises the optical depth to $\tau_{10,cor}=1.12\pm0.04$. However, the value of $C_f$ is relatively uncertain [e.g., \citet{Henkel08}], and a slight difference in the covering factor could have a significant impact. For instance, 
$C_f=0.3$ and $C_f=0.4$ would imply $\tau_{10,cor}=1.54\pm0.07$ and $\tau_{10,cor}=0.89\pm0.03$, respectively. For the non-lensed systems, covering factors may also be applicable. For 
AO 0235+164 we use $C_f=1.0$ \citep{Curran05,Kanekar09}. For the Q0852+3435, Q0937+5628, and Q1203+0634 QASs estimates of the 
covering factor are not available in the literature. Although \citet{Kanekar09} find a median covering factor of $C_f=0.885$ for a sample of z$<$0.9 DLAs, this value was determined at low frequencies
($\lesssim$ 1.1~GHz) where extended radio structure may be contributing to the emission. Since such extended structure is unlikely to be associated with the silicate dust, we instead conservatively adopt a value of 
$C_f=1.0$ for all QASs without a published covering factor. 
For the z=0.437 QAS toward 3C~196, although the 21 cm absorption has a very low covering factor against the AGN [$C_f \lesssim 0.1$; \citet{Kanekar09}], 
it is not clear that such a low covering factor is appropriate for the dusty QAS material, due to the differences between the radio and optical sightlines to the quasar.
Thus, we again adopt a representative $C_f=1.0$ for the 3C~196 QAS. Given the significant 
uncertainties and assumptions associated with some of these covering factors, when examining correlations with other dust and gas properties we utilize both the measured and the corrected 
peak optical depth measurements, rather than relying solely on the covering factor-corrected values.

\begin{deluxetable}{llllllllll}
\tabletypesize{\scriptsize}
\rotate
\tablecaption{Comparison with Other Program QASs}
\tablewidth{0pt}
\tablehead{
\colhead{QAS} & \colhead{z$_{abs}$} & \colhead{Mineral} & \colhead{$\tau_{10,obs}$} &  \colhead{$C_f$} & \colhead{$\tau_{10,cor}$} & \colhead{E(B-V)} & \colhead{EW$_{2796}$} & \colhead{$\tau_{21}$}& \colhead{$\tau_{21,cor}$}}
\startdata
AO 0235+164 & 0.524 & Amorph.Oliv.GS (1)& 0.086$\pm$0.003\tablenotemark{a}~(2) & 1.0~(3,4) & 0.086$\pm$0.003 & 0.227$\pm$0.003 (5) & 2.42$\pm$0.06 (6) & $\geq$0.71 (7) &0.71\\ 
3C 196 & 0.437 & Amorph.Oliv.2.0 (8) & 0.103$\pm$0.007\tablenotemark{a,b}~(2) & 1.0\tablenotemark{c}& 0.103$\pm$0.007 &0.20$\pm$0.07 (2,9) & 2.00$\pm$0.07 (10) & 0.05 (2,11)&\nodata\\ 
Q0852+3435 & 1.309 & Amorph.Oliv.GPC (1)  &0.165$\pm$0.010\tablenotemark{a}~(2) &  1.0\tablenotemark{c}& 0.165$\pm$0.010 &0.40$\pm$0.12 (2)& 2.97$\pm$0.12 (2) &0.14 (2,12)&0.14\\  
Q0937+5628 & 0.978 & Amorph.Oliv.GPC (1) & 0.192$^{+0.018}_{-0.017}$\tablenotemark{a}~(2) & 1.0\tablenotemark{c} &0.192$^{+0.018}_{-0.017}$ &0.34$\pm$0.12 (2) & 5.77$\pm$0.22 (2) & \nodata&\nodata\\ 
Q1203+0634 & 0.862 & Amorph.Pyrox (13) & 0.277$^{+0.041}_{-0.040}$\tablenotemark{a,d}~(2) &1.0\tablenotemark{c} & 0.277$^{+0.041}_{-0.040}$ &0.54$\pm$0.18 (2) & 5.52$\pm$0.30 (2)&\nodata &\nodata\\ 
PKS 1830-211 & 0.886 & Hortonolite (14) & 0.269$\pm$0.006\tablenotemark{e}~(15) & 0.35~(16,17) & 1.120$^{+0.041}_{-0.039}$ &0.57$\pm$0.13(18)&\nodata& 0.055 (19) &0.17\\ 
TXS 0218+357 & 0.685 & Amorph.Oliv.GPC (1) & 0.493$\pm$0.024 & 0.8~(20,21) & 0.667$^{+0.036}_{-0.035}$ &0.62$\pm$0.04~(18)&3.0 (22) &0.048 (23) &0.06\\ 
\enddata
\tablenotetext{a}{Quasar continuum normalized spectrum from \citet{Kulkarni11} re-fitted using methodology described in \S\ref{results}.} 
\tablenotetext{b}{If a power-law quasar continuum normalization were adopted, a shallower optical depth (as low as $\tau_{10}\sim0.05$) could be obtained, depending on the adopted silicate template.}
\tablenotetext{c}{For QASs with no measurements of the covering factor in the literature, a value of 1.0 is assumed.}
\tablenotetext{d}{The fit using the Am.Oliv. template described in \citet{Kulkarni11} has a consistent peak optical depth, but slightly poorer fit-quality.}
\tablenotetext{e}{Uncertainty recalculated to match methodology described in \S\ref{results}.}
\tablecomments{Comparison of the silicate absorption in the 7 studied QASs (column 1). We list the (column 2) QAS redshift; 
(column 3) laboratory mineral chosen from \citet{Aller12} Table 10 list which produces the best fit to the 10~$\micron$ feature (and reference); (column 4) measured peak 10~$\micron$ silicate optical depth; 
(column 5) adopted covering factor (and reference); (column 6) peak 10~$\micron$ silicate optical depth corrected using adopted covering factor; (column 7) E(B-V) reddening; 
(column 8) QAS rest-frame equivalent width for the Mg~II $\lambda$2796 absorption (in \AA); 
(column 9) the 21-cm optical depth; and (column 10) the 21-cm optical depth corrected using the adopted covering factor. 
The reddening, Mg~II equivalent width, and 21-cm optical depth are obtained from the literature, with the reference given in parenthesis. 
When multiple reddening values were presented in a single analysis, the value with the lowest $\chi^2$ was adopted. 
References: (1) \citet{Chiar06} based on data from \citet{Henning99}; (2) \citet{Kulkarni11}; (3) \citet{Curran05}; (4) \citet{Kanekar09}; (5) \citet{Junkar04}; (6) \citet{Wolfe77}; (7) \citet{Roberts76}; 
(8) \citet{Dorschner95}; (9) \citet{Gharanfoli07}; (10) \citet{Foltz88}; (11) \citet{Briggs01};  (12) \citet{Srianand08};  
(13)  \citet{Spoon} based on data from \citet{Fabian01}; (14) \citet{Jaeger} with tabulated data provided by C. J\"{a}ger;  
(15) \citet{Aller12}; (16) \citet{Frye97}; (17) \citet{Henkel08};  (18) \citet{Falco99};  (19) \citet{Chengalur99}; (20) \citet{Wiklind95}; (21) \citet{Muller07}; (22) \citet{Browne93}; \& (23) \citet{Carilli93}}\label{COMPARE} 
\end{deluxetable}
 
\subsection{10-to-18~$\micron$ Silicate Dust Optical Depth Ratio}
Silicate dust grains in the ISM produce 2 prominent spectral features, which we see in the TXS 0218+357 QAS: a 10~$\micron$ and an 18~$\micron$ feature. 
The former is produced from an Si-O stretching mode in the dust molecules, while the latter originates in an O-Si-O bending mode. The amount of absorption 
produced by these two features, i.e. the ratio of their peak optical depths, varies with changes in dust grain properties such as the 
distribution of grain shapes and sizes. Analyses of the ISM in Local Group galaxies [e.g., \citet{Weingartner01}], as well as those in higher redshift systems such as AGN tori [e.g., \citet{Hao05}], 
have utilized the observed ratio of these features to constrain dust grain properties. 

In the TXS 0218+357 QAS, we find the peak optical depth ratio calculated from the observed flux ratios to be $\tau_{10}/\tau_{18}= 1.31\pm0.48$. 
The best-fitting (lowest $\chi_r^2$) amorphous olivine template, which had porous, ellipsoidal particles 
\citep{Chiar06}, has a $\tau_{10}/\tau_{18}$ ratio of 1.41, which more closely matches the data than, for example, a similar amorphous olivine template with solid, 
spherical particles that predicts a $\tau_{10}/\tau_{18}$ ratio of 1.62. 
(However, it is not possible to definitively rule out solid, spherical particles given the large uncertainty in the  $\tau_{10}/\tau_{18}$ ratio inferred from the data.)
While in principle,
we could extend our analysis to compare our observed silicate dust opacity curves with models constructed for different grain temperatures and geometries 
[e.g. \citet{Li01,Weingartner01}], we have not taken this step because of the low S/N of our data, and the uncertainties associated with our quasar continuum normalization. As discussed in Appendix~\ref{normAP},
since our data do not extend past 21.5~$\micron$ in the QAS rest-frame, the shape of the quasar continuum underlying the 18~$\micron$ silicate feature is poorly constrained. Differing
assumptions about the physical shape of the quasar continuum can significantly affect the depth of the 18~$\micron$ feature. We, therefore, feel that a rigorous comparison with the dust grain models is
not warranted for this dataset. 

\subsection{Connection between Reddening and Silicate Absorption}

Given the high peak optical depth measurement for the silicate dust in the TXS 0218+357 QAS relative to the 6 previously studied QASs, we have revisited the suggested trend
between silicate absorption peak optical depth and reddening presented in \citet{Kulkarni11}. In that work, it was noted that the trend seemed similar
to that for Milky Way diffuse ISM clouds, i.e. $\tau_{10}=0.17 \times E(B-V)$ for R$_V$=3.1, but that the slope differed such that the ratio of $\tau_{10}/E(B-V)$ was 2-4 times that of diffuse Galactic clouds. 
This was used to argue for either a greater abundance of silicate dust in these QASs than is found in the Milky Way ISM dust, possibly stemming from differences
in the underlying stellar populations, or for differences in the dust grain geometrical and size properties which would affect R$_V$. 

\begin{figure} 
\epsscale{1.0} 
\plottwo{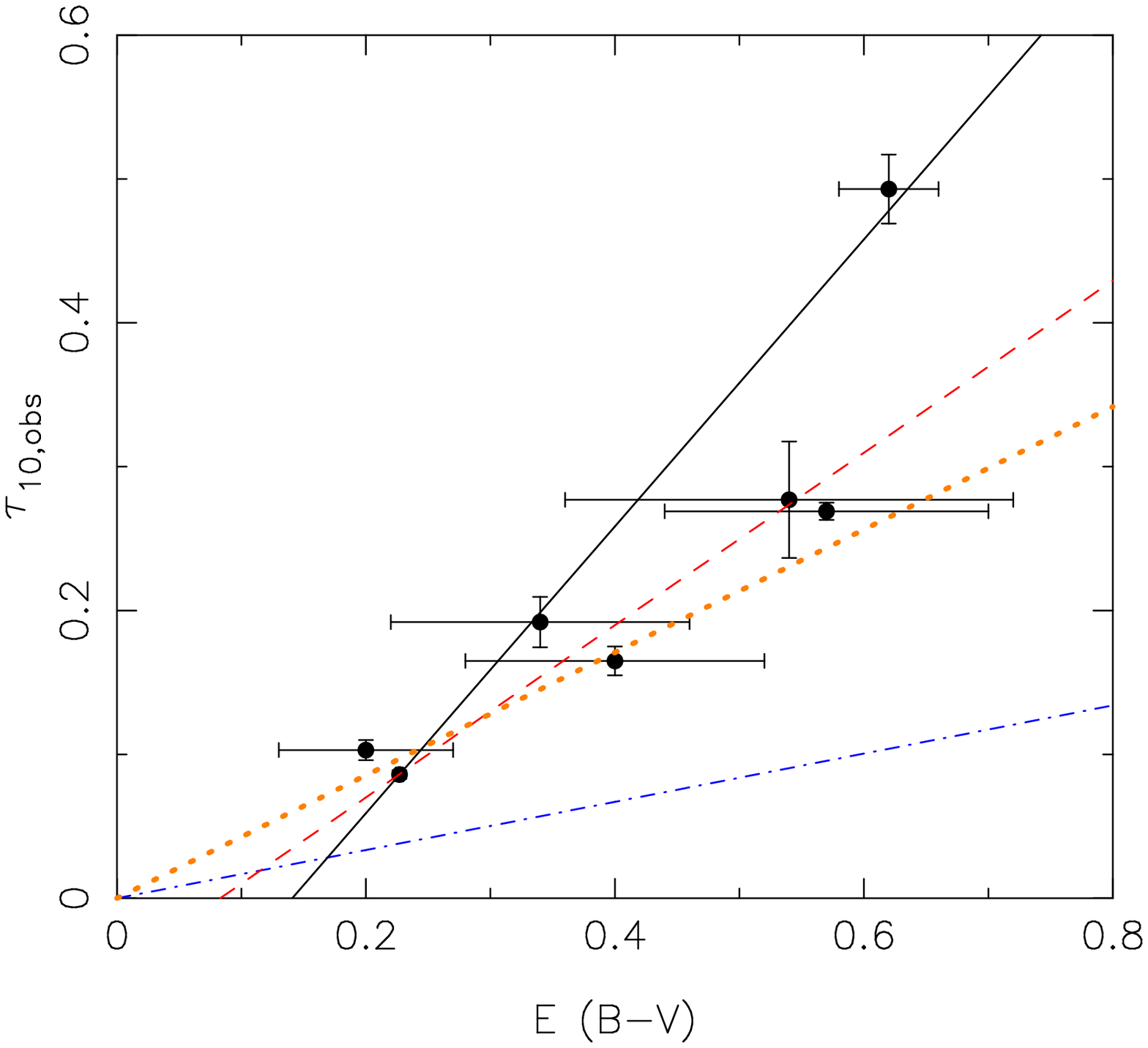}{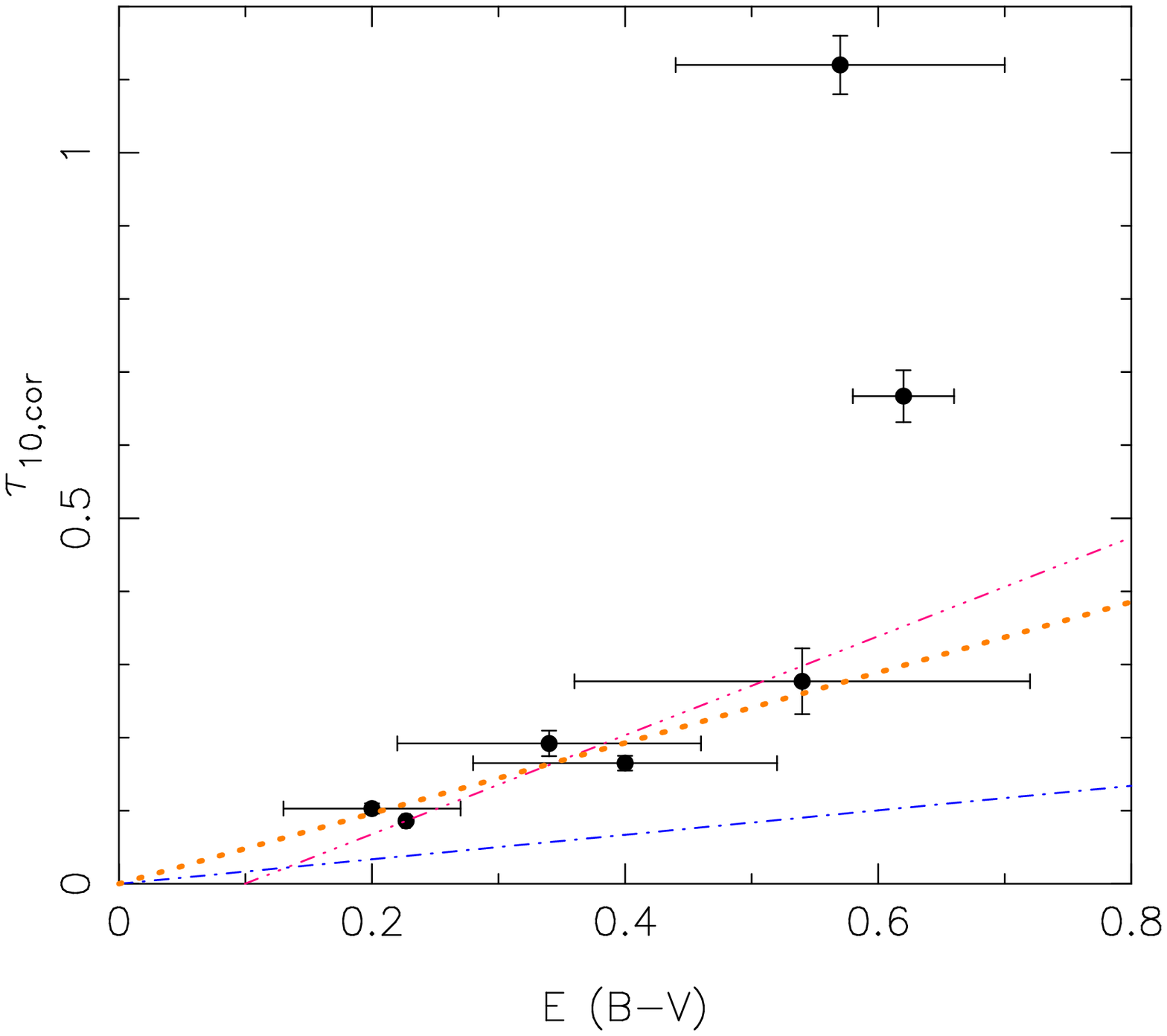}%
\caption{Illustration of possible correlation between the silicate peak optical depth and the reddening.  
The left panel utilizes the measured 10~$\micron$ peak optical depths, while in the right panel the optical depths are corrected using the covering factor.
In the left panel, the solid black line shows the linear least-squares fit to all 7 QASs, and the red dashed line shows the fit which excludes the TXS 0218+357 absorber.
In the right panel, the magenta dash-dot-dot-dotted line shows the fit excluding both the TXS 0218+357 and PKS 1830-211 QASs. 
In both panels, the orange dotted line shows the fit required to pass through the origin.
The shallow blue dot-dashed line is an extrapolation of the trend for Milky Way diffuse clouds from \citet{Roche84}.  
[See the electronic edition of the Journal for a color version of this figure.]} \label{Ebvplot} 
\end{figure}

The addition of the TXS 0218+357 QAS and the z=0.886 PKS 1830-211 QAS to the 5 previously studied QASs reveals a clear trend of increasing reddening with increasing silicate
measured peak optical depth, as illustrated in Figure~\ref{Ebvplot} (left). The total reddening measurements are from the literature, as detailed in Table~\ref{COMPARE}. 
Considering all 7 QAS we find a Pearson linear correlation coefficient of 0.89 (0.8\% probability of no correlation), a Spearman (nonparametric) rank correlation coefficient of 0.89 
(significance 0.007, i.e., 0.7\% probability of no correlation), and a (nonparametric) Kendall's Tau value of 0.71 (significance 0.02). 
The small probabilities/significances indicate a relatively strong correlation. 
We fitted a linear least-squares relationship to the 7 QASs, weighting by both the uncertainties in the reddening from the literature and by those for our measured peak silicate optical depths. 
We find a slope of 1.00$\pm$0.11, which is substantially steeper than the 0.17 slope found for the extrapolation for Milky Way diffuse clouds from \citet{Roche84}. 
Our fit is illustrated in Figure~\ref{Ebvplot} (left) with a solid black line, while the shallower Milky Way extrapolation is shown as a blue dot-dashed line. 
All of the studied QASs lie significantly above the Milky Way extrapolation. We note that the steepness of the slope we obtain is being driven in part
by the TXS 0218+357 QAS. However, even excluding that QAS gives a slope of 0.60$\pm$0.18 (dashed red line), which is much steeper than the Milky Way extrapolation.
If we require our relationship to extrapolate as a smooth linear function to dust-free systems, i.e. to predict an optical depth of 0.0 for no reddening, we would get a still relatively steep slope  
of 0.43$\pm$0.01 (shown as dotted orange line in Figure~\ref{Ebvplot}). While this slope is closer to that predicted from the Milky Way, it remains nearly three times higher than the slope for the extrapolated Milky Way relation, and 
the TXS 0218+357 QAS would lie significantly above the relationship. Since the reddening error bars are sensitive to the adopted intrinsic quasar spectrum in the extinction-curve-fitting reddening 
measurements [e.g., \citet{Kulkarni11}] we have also examined
the slope obtained if we omit the reddening uncertainties when fitting over all 7 QASs. We find a slope of 0.51$\pm$0.02, which does not alter our conclusions.

We have also explored the relationship between the reddening and the silicate peak optical depth using the covering factor-corrected peak optical depths, as depicted in 
Figure~\ref{Ebvplot} (right). Since the covering factors for several of the QASs are highly uncertain, we place less confidence in the derived slope of the relationship, although there 
is still a general trend of increasing silicate peak optical depth with increasing reddening.
Furthermore, since the optical depth values are increased by applying the covering factor correction, the offset relative to the Milky Way diffuse cloud extrapolation increases for $C_f<1$. 
The Pearson correlation coefficient is 0.76 (5\% probability of no correlation), with a Spearman (nonparametric) rank correlation coefficient of 0.89 
(significance 0.007), and a (nonparametric) Kendall's Tau value of 0.71 (significance 0.02). 
The magenta dash-dot-dot-dotted line depicts the linear least-squares fit (slope of 0.68$\pm$0.31) obtained when excluding the TXS 0218+357 and PKS 1830-211 QASs. 
When corrected for the covering factors, the molecule-rich PKS 1830-211 and TXS 0218+357 QASs lie above the trend suggested by the other QASs, i.e., their
silicate peak optical depths are higher than expected given their reddening values. This is particularly interesting in the context of studies in the Milky Way [e.g., \citet{Chiar07}], which find that while
there is a tight correlation between the 10~$\micron$ silicate peak optical depth and the near-infrared color excess $E(J-K)$ along diffuse ISM sightlines, when molecular clouds
are examined the correlation is absent, with molecular sightlines exhibiting \textit{lower} silicate optical depths for a given $E(J-K)$. The reason for this
difference between diffuse and molecular cloud sightlines is still not completely clear. It has been argued that the origin is an increase in $E(J-K)$ in molecular clouds, 
possibly linked to grain growth or variations in dust species abundances \citep{vanBreemen11}. Our molecule-rich systems, however, lie above the trend established
by the other QASs. This is more similar to the Galactic center sightline, which exhibits a higher silicate optical depth for its observed $E(J-K)$ relative to other Galactic diffuse ISM sightlines, possibly because of fewer C-rich stars near the Galactic center. \citep{Roche85}.

Overall, the disparity between the Milky Way extrapolation and the studied QASs seems most significant in systems with large silicate optical depths; these systems drive the steeper derived slope relative to the Milky Way. 
This may indicate that in distant galaxies at relatively high silicate dust columns, the selective extinction differs from that in the Milky Way because of differing dust grain properties. 
One possible explanation is that the interstellar dust grains may be larger than those in the Milky Way ISM. Large dust grains would yield a low UV extinction, which would 
translate into a high R$_V$, consistent with the measured R$_V\sim$7.2 for the TXS 0218+357 QAS \citep{Falco99}. This decrease in UV extinction, which is not accompanied by a similar decrease
in IR absorption, could produce the offset relative to the Milky Way extrapolation. Furthermore, we note that reddening measurements in the Milky Way sample an integrated combination of different dust
grain populations along the Galactic plane sightline, while systems such as the TXS 0218+357 QAS are viewed nearly face-on \citep{York05}. 
This difference in viewing angle may mean that a single, less common, type of dust grain
could disproportionately impact the derived total reddening for face-on galaxies.
However, the good agreement between our silicate absorption profile and that for amorphous olivine,
which has also been used to describe Milky Way ISM silicate dust grains, would imply that the silicate dust grains are somewhat similar. 
This might suggest that the disparity stems instead from a difference in the underlying stellar populations producing the dust.
A larger sample of QASs, particularly QASs with large silicate optical depths, are needed to more strongly constrain the slope of the reddening-silicate optical depth relationship,
and to investigate whether the relationship is adequately described by a simple linear trend. Furthermore, measurements of the silicate optical depth in 
QASs with low reddening values, in which the silicate optical depth is expected to be low requiring (future) high-sensitivity instruments, will help to establish
whether the trend extends to systems with less dust. 

\subsection{21-cm Absorption in TXS 0218+357}
\begin{figure} 
\epsscale{1.0} 
\plottwo{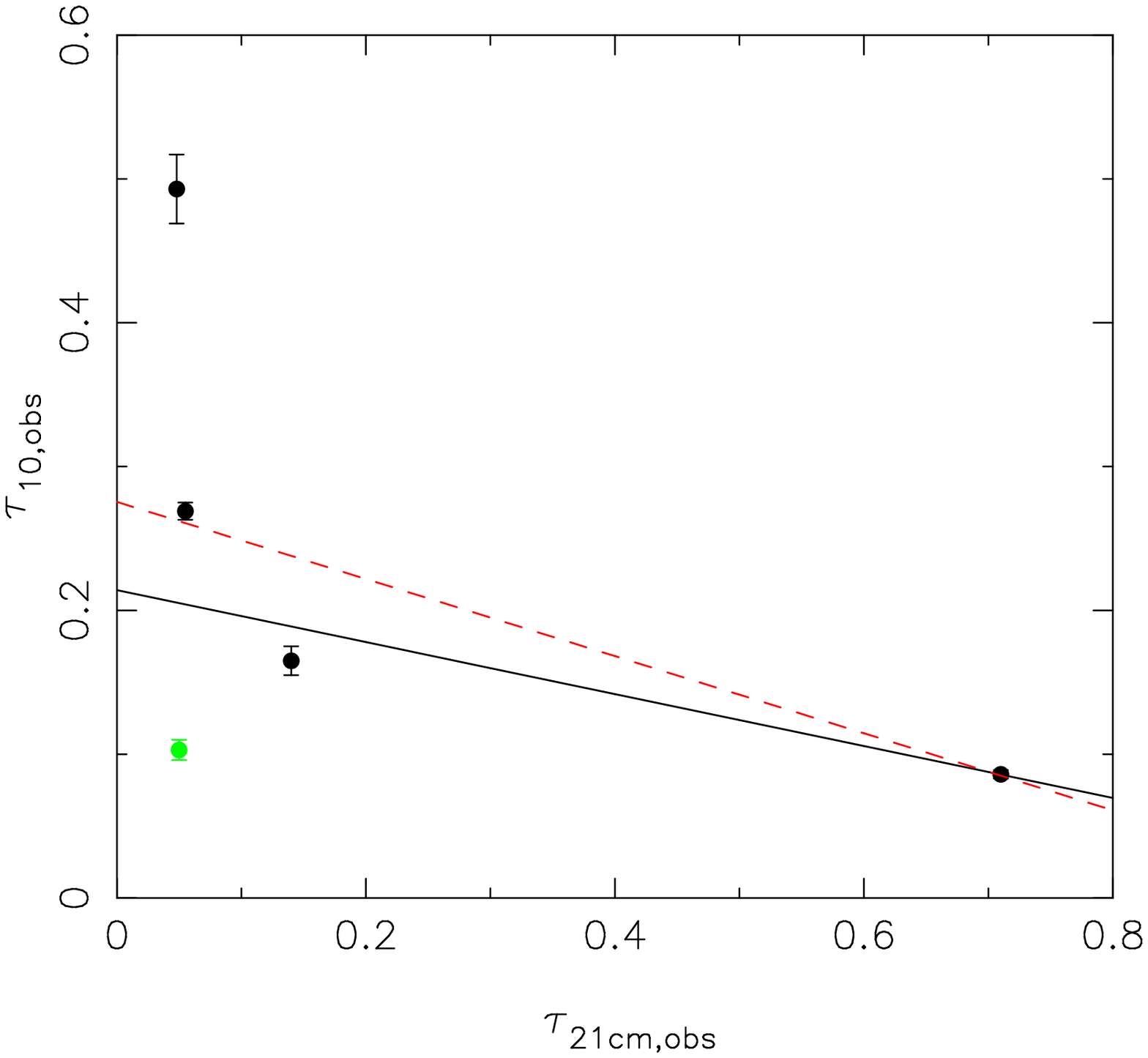}{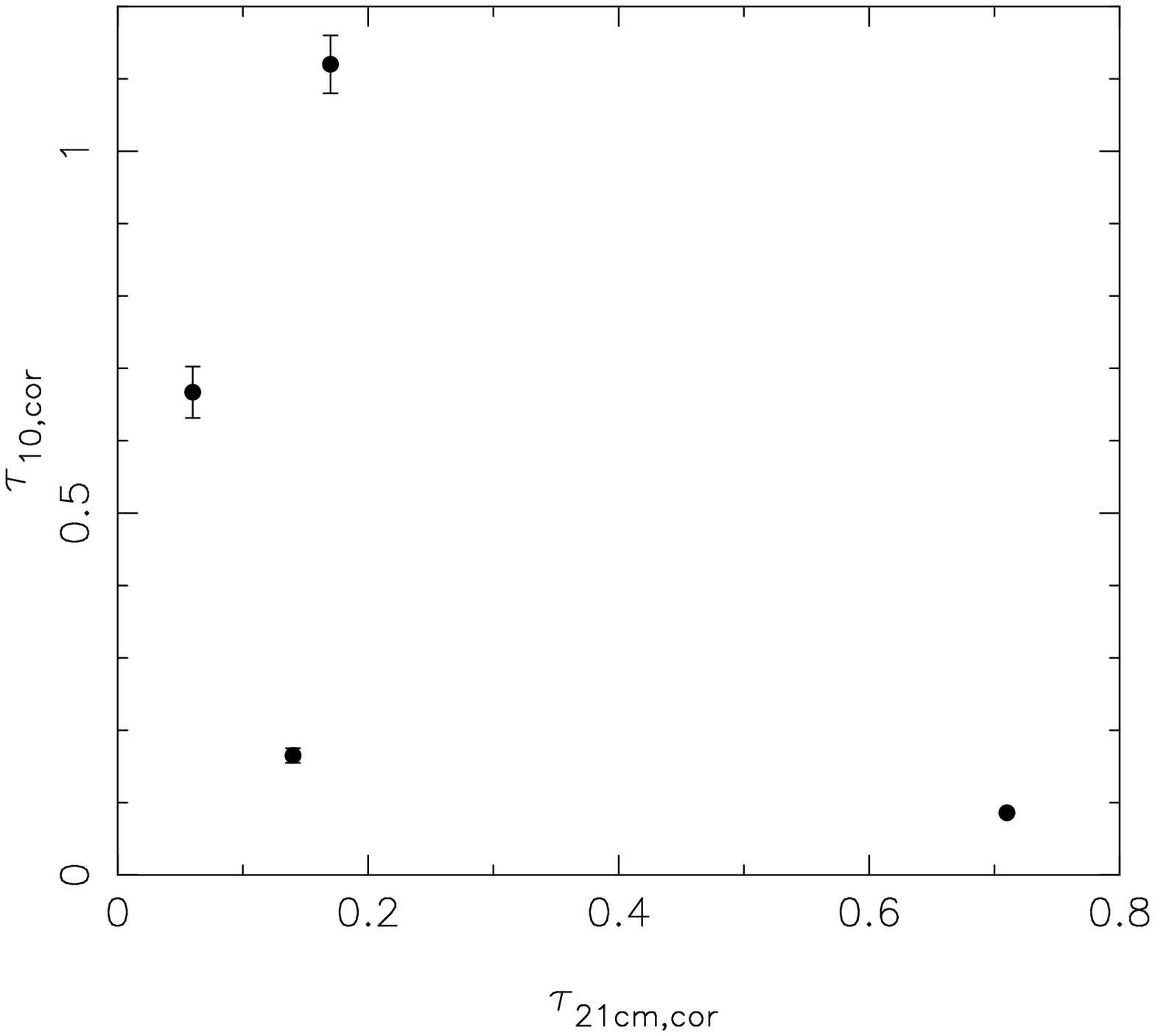}%
\caption{21 cm H~I optical depth plotted versus the silicate peak optical depth for the 5 QASs with published 21 cm measurements.  
The left panel utilizes the measured peak optical depths, while in the right panel the optical depths are corrected using the covering factor.
The 3C 196 QAS is depicted in green in the left panel, and excluded in the right panel, since the optical and radio sightlines may differ in this system, and a reliable estimate of the covering factor is not available for this QAS. 
In the left panel, the black solid line shows a linear fit using all 5 QASs, while the red dashed line excludes the 3C 196 QAS measurement from the fit.  
There appears to be a suggestion of a slight anti-correlation between the 10~$\micron$ and 21 cm peak optical depths. 
[See the electronic edition of the Journal for a color version of this figure.]} \label{T21plot} 
\end{figure}

The 21-cm peak optical depth provides an estimate of the absorption produced by the cold, dusty H~I gas in the ISM. 
We list measurements of the observed $\tau_{21cm}$, taken from the literature, in Table~\ref{COMPARE}. 
In the TXS 0218+357 system, \citet{Carilli93} find a peak optical depth of $\tau_{21cm}=0.048\pm0.005$, assuming a covering factor of 1.0. The line structure is noted
to be somewhat asymmetrical, but under an assumed Gaussian fit a FWHM of 43$\pm$4 km/s is derived. The velocity width, optical depth, and shape for this 21-cm absorption feature 
are all noted by  \citet{Carilli93} to be similar to those seen in the z=0.437 QAS toward 3C~196, although concerns
that the 3C~196 absorption is occurring against the lobes rather than against the AGN core imply that they may not be directly comparable \citep{Briggs01}. This sightline difference 
may explain why despite the similarity in their measured 21-cm peak optical depths, the measured peak silicate optical depths differ:
$\tau_{10}$ is nearly five times larger for the TXS 0218+357 QAS than for the z=0.437 QAS toward 3C~196. 
The remaining systems with 21cm absorption measurements suggest that there may be a slight anti-correlation between the measured silicate peak optical depth and the measured 
21 cm peak optical depth, as depicted in Figure~\ref{T21plot} (left). However, as evidenced by the poor match of the linear least-squares fit to the data, this may not 
be a purely linear relationship, but more data points are required to establish the precise shape of the relationship. When the covering factors are applied to the 21 cm and silicate peak optical  
depth measurements (Figure~\ref{T21plot}, left), the trend becomes less clear, largely because of the high derived $\tau_{10,cor}$ optical depth for the PKS 1830-211 system, which has a low covering factor. 
Although a slight anti-correlation is suggested by the data, a larger data sample is required to investigate further. In any case, variations in properties of the material, such as the dust-to-gas ratio and the depletions, could
induce scatter in even an intrinsically tight correlation. 

\subsection{Correlation between Mg~II Equivalent Width and Silicate Dust}
We also investigate suggestions of a correlation between the Mg~II $\lambda$2796 rest-frame equivalent width and the peak optical depth of the silicate dust absorption. 
\citet{Kulkarni11} found a correlation between these properties and suggested that the more silicate-rich QASs may be more massive. This is because since the Mg~II lines are 
saturated in the absorbers studied by \citet{Kulkarni11}, the Mg~II equivalent width is a better proxy for the velocity
spread (i.e. the absorber mass or presence of outflows into the CGM) than for the elemental abundance of metals in the ISM gas. 

\begin{figure} 
\epsscale{1.0} 
\plottwo{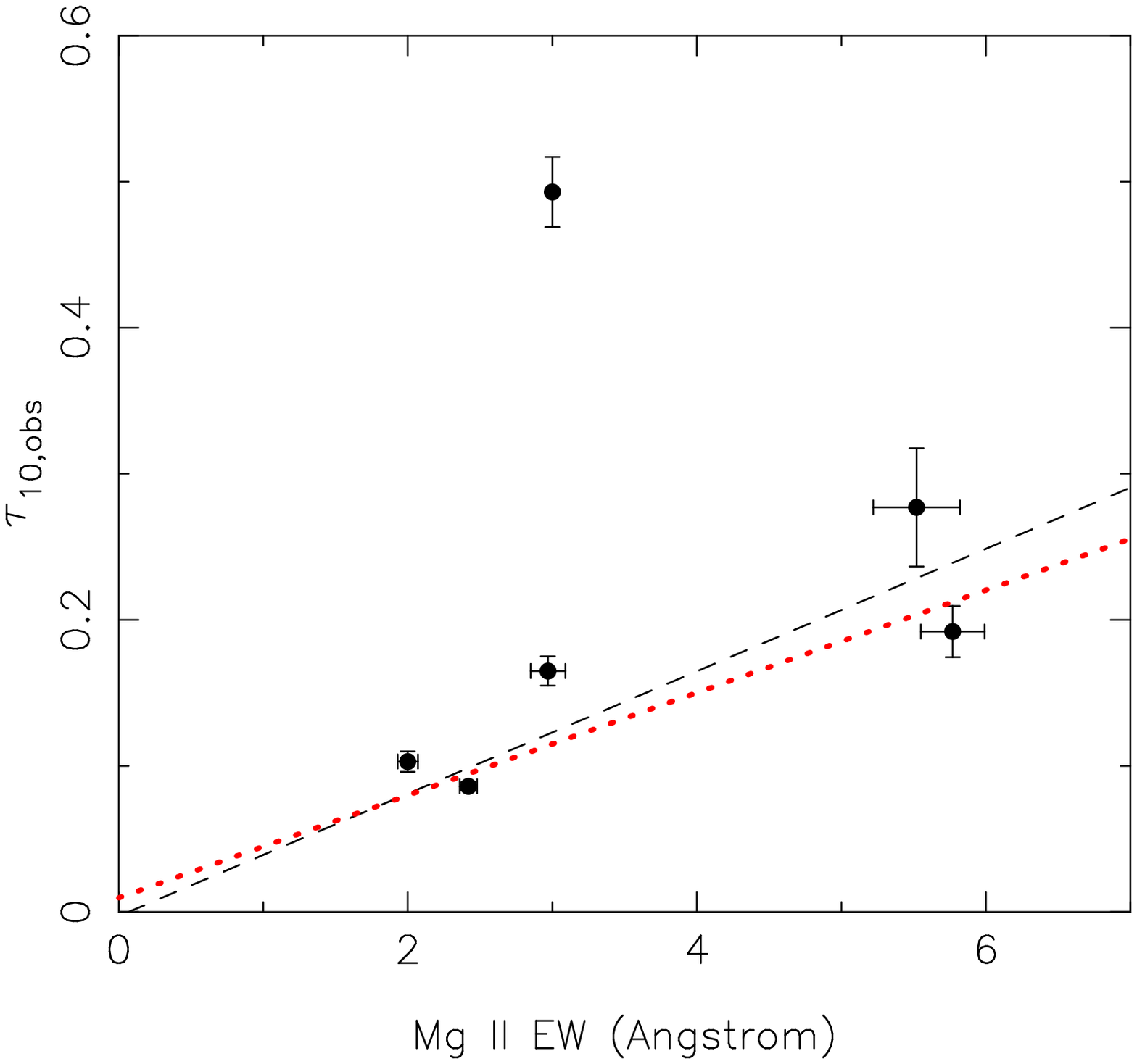}{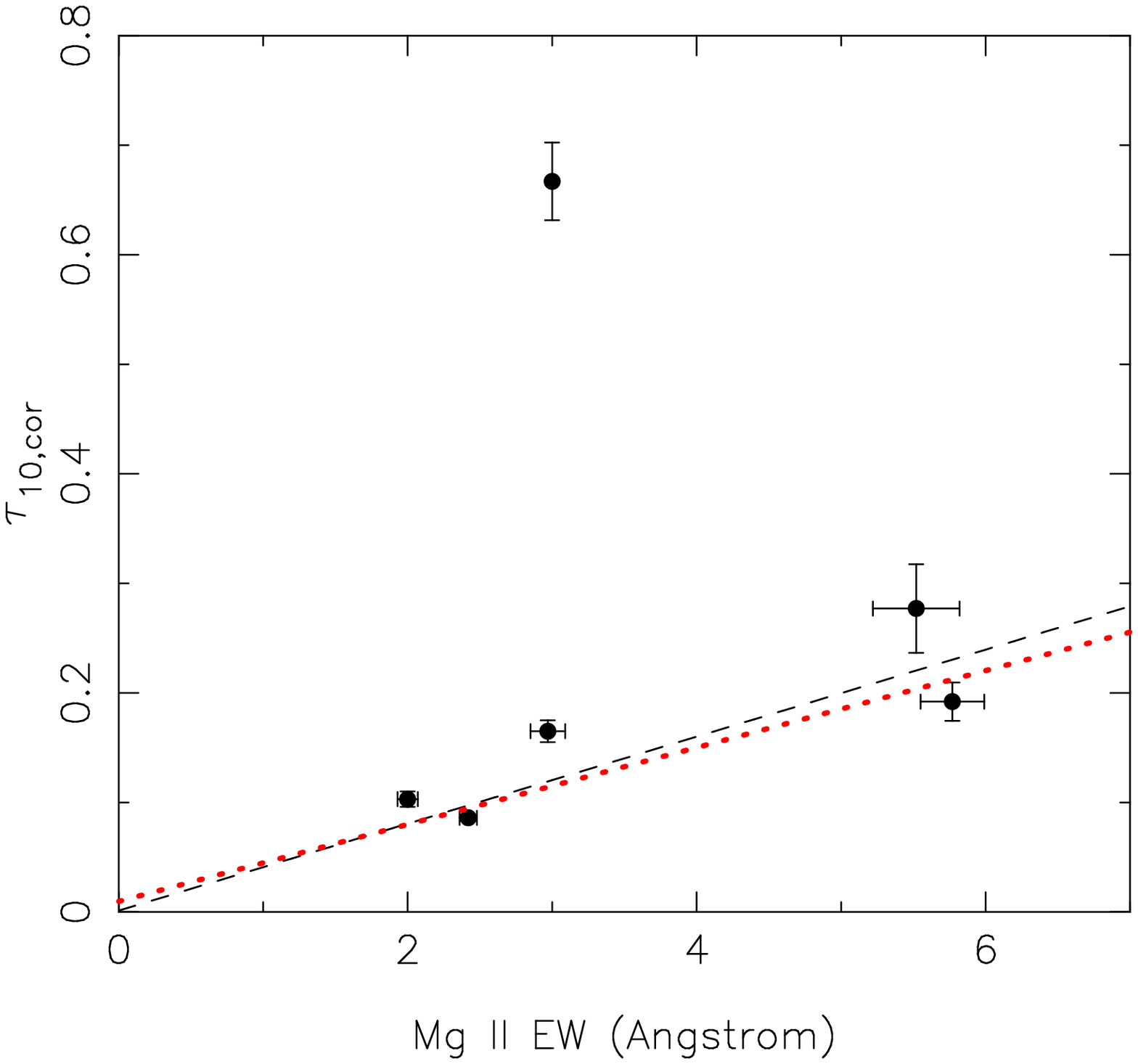}%
\caption{Relation between the rest-fame Mg~II $\lambda$2796 equivalent width and silicate peak optical depth for the 6 QASs with published Mg~II measurements.  
The left panel utilizes the measured 10~$\micron$ peak optical depths, while in the right panel the optical depths are corrected using the covering factor. 
The black dashed line shows the fit using all 6 QASs, while the red dotted line depicts the fit excluding the TXS 0218+357 QAS. 
The TXS 0218+357 QAS appears to be an outlier compared to the relationship between these parameters for the other program QASs.  
[See the electronic edition of the Journal for a color version of this figure.]} \label{Mgplot} 
\end{figure}

We have revisited the  correlation here with the addition of the TXS 0218+357 QAS, as shown in Figure~\ref{Mgplot} using both the measured (left) and corrected (right) silicate peak optical depths. 
However, we find that the TXS 0218+357 QAS is an outlier relative to the other 5 QASs, with a much stronger peak optical depth than would be predicted from its Mg~II equivalent width.
Including all 6 QASs and using the measured peak optical depths, these parameters have a Pearson correlation coefficient of 0.22 (68\% probability of no correlation), a Spearman correlation coefficient of 0.71  
(significance 0.11), and a Kendall's Tau value of 0.47 (significance 0.19).  
If we exclude the TXS 0218+357 outlier point, however, we would obtain a stronger Pearson correlation coefficient of 0.86 (6\% probability of no correlation) and Kendall's Tau of 0.60 (significance 0.14). 
A linear least-squares fit to all 6 QASs, weighting solely by the $\tau_{10}$ uncertainties, since measurement uncertainties are not available for all of the Mg~II measurements in the literature,
produces a slope of 0.042$\pm$0.005 for the measured optical depth values, and 0.040$\pm$0.005 using the covering-factor corrected optical depths. 
This relationship is illustrated as a dashed line. The previously studied 5 QASs cluster around this line, while the TXS 0218+357 
QAS lies substantially above the fit.  If the TXS 0218+357 QAS is excluded from the fit, the slope decreases slightly to 0.035$\pm$0.005 (same for the corrected optical depths), as depicted with the dotted line. 
A larger sample will be required to see whether the TXS 0218+357 QAS is anomalous, or whether the suggested correlation is not as tight was was implied with the earlier sample of QASs. 

Some of the discrepancy for the TXS 0218+357 QAS may arise if the Mg~II line in this system is not saturated, unlike in the 5 previously studied QASs.
This may be caused in part by the Mg~II slit encompassing both quasar sightlines, one of which has significantly less absorption and is brighter at optical wavelengths.
While the equivalent width ratio of the lines is suggestive of saturation, the resolution and S/N of the spectrum are not high enough to definitively assess if the lines are saturated \citep{Browne93}. 
If the lines are unsaturated, the Mg~II measurement may be providing an estimate more reflective of the elemental abundance than of the galaxy dynamics (i.e. the velocity spread). 
Alternatively, it is possible that the sampled sightline through the TXS 0218+357 QAS's nearly face-on galaxy probes inner regions which produce less efficient galactic outflows of warm gas containing Mg~II, and that
if viewed from a different sightline a higher Mg~II equivalent width would be obtained. 
If an elemental discrepancy is invoked,
the ISM gas in this system may be Mg-poor either because of an intrinsic deficit in the total (gas+solid phase) amount of Mg, which would also result in less Mg-rich (more Fe-rich) species of olivines/pyroxene silicates.
Alternatively, the gas may be Mg-poor because of depletions onto the dust grains, resulting in Mg-rich olivines/pyroxenes. Data with higher spectral resolution and S/N, and with longer wavelength coverage, 
are required to discriminate better between these silicate species. 

\subsection{Connections Between Silicate Crystallinity and Molecule Formation?}
Lastly, the TXS 0218+357 QAS provides an interesting opportunity to investigate whether there is a link between silicate crystallinity and a high enrichment of molecular material. 
The degree of silicate crystallinity depends on the conditions (e.g., temperature) under which the dust grains formed and the subsequent processing of the dust grains (e.g., by cosmic rays, grain-grain collisions), 
and, therefore, has implications for the evolution of the dust grains [e.g., \citet{Kemper11}].
In the z=0.886 QAS toward PKS 1830-211, a system which has been noted to have the largest number (28) of detected molecular species in any extragalactic object to date \citep{Muller11},
we previously found very suggestive evidence for a high degree of silicate crystallinity \citep{Aller12}. This evidence for crystallinity stems from the shape and location of the substructure in the 
observed QAS 10~$\micron$ silicate absorption feature. Since in the Milky Way ISM the silicate dust is predominately amorphous \citep{Kemper04, Li07}, this raises the question of how the
silicate dust in PKS 1830-211 could be so much more crystalline, and whether it is linked to the high degree of molecule-enrichment. For instance, does some aspect (e.g. temperature, density)
of the ISM environment facilitate the formation of both molecules and crystalline silicate grains? Since the TXS 0218+357 QAS is also noted to be very rich in 
molecules, and exhibits 10~$\micron$ silicate absorption, it provides a unique opportunity to test this hypothesis. 

The TXS 0218+357 QAS has been noted to exhibit signatures of many molecules, 
including CO, HCO$^+$, HCN, HNC, CN, CS, C$^{18}$O, $^{13}$CO, H$_2$O, NH$_3$, OH, H$_2$CO, and LiH 
 \citep{Wiklind95, Combes95, Combes97b, Combes98,Combes97,Gerin97,Menten96,Kanekar02,Kanekar03,Henkel05,Jethava07,Zeiger10,Friedel11}. 
As for the lensed PKS 1830-211 z=0.886 QAS, it appears that one of the two quasar sightlines (both of which are encompassed by the \textit{IRS} slit) is more enriched.
Although a direct comparison of the abundances in the z=0.886 PKS 1830-211 QAS and the TXS 0218+357 QAS is challenging, since derived molecular abundances depend on assumptions 
about the degree of line saturation, the covering factor, and the excitation temperature, for most molecules it appears that the TXS 0218+357 QAS is slightly less enriched. 
The OH and HCO$^+$ column densities are both good predictors of N$_{H_2}$ in Galactic clouds \citep{Liszt99}. 
A direct comparison in \citet{Kanekar02} finds that the derived N$_{OH}$ and N$_{HCO+}$ column densities in TXS 0218+357 are 19-23\% of those for the z=0.886 PKS 1830-211 absorber.
Using OH as a tracer of the molecular hydrogen results in a column density of N$_{H_2}$ of $2.65\times 10^{22}$ cm$^{-2}$ for the TXS 0218+357 \citep{Kanekar02}. 

However, while the evidence for significant molecule enrichment in the ISM in the TXS 0218+357 QAS is undisputed, the evidence for or against silicate crystallinity is much less clear due to the low S/N of our 
data. The combination of a lower flux density, fewer exposures, and a smaller total exposure time for the TXS 0218+357 quasar, in comparison with PKS 1830-211 \citep{Aller12}, makes it much more difficult to establish the presence or absence
of silicate crystallinity. As discussed in \S\ref{results} the laboratory amorphous olivine laboratory template for the TXS 0218+357 system provides a good fit, both in terms of the breadth and depth of the 10 and 18~$\micron$
silicate absorption features, and in terms of the peak absorption wavelength. However, as illustrated in Figure~\ref{SINGLEFIT}, the magnitude of the substructure in the 10 and 18~$\micron$ regions for some species of crystalline olivine
and pyroxene are small enough to be masked by the large uncertainties and noise associated with our spectrum. Several of the predicted substucture features align relatively well with the possibly noise features present in our data. 
Furthermore, as established with the two-mineral fits in \S\ref{twofitsect}, a combination of crystalline olivine and pyroxene would produce an excellent match to the 10~$\micron$ and 18~$\micron$ silicate absorption features; 
the predicted abundant substructure would occur at levels below that of the noise in the data. Thus, although the smooth, featureless amorphous olivine fit is both consistent with the ISM seen in the Milky Way and with our data, 
we cannot rule out that abundant silicate crystallinity \textit{could} be present in the TXS 0218+357, although it seems less likely than in the PKS 1830-211 z=0.886 QAS. 

If the explanation that the TXS 0218+357 QAS is produced by amorphous olivines is adopted, it establishes a similarity between the silicate dust in this z=0.685 galaxy and that  in the Milky Way ISM. 
By contrast, independent of the substructure, the redder peak wavelength of the absorption feature in the PKS 1830-211 z=0.886 QAS implies 
silicate dust grain differences with that in the Milky Way ISM and with that in the TXS 0218+357 QAS ISM. In combination, these data suggest that there
could be significant variations in the dust grain properties between different galaxies, which may not be a simple function of redshift evolution or of molecular gas abundance. 
Further investigation of the silicate dust grains in the TXS 0218+357 with higher spectral resolution and S/N data, as well as a study of a broader sample of QASs, is required to quantify these variations and establish whether
ISM silicate dust grain crystallinity varies between galaxies as a function of composition, star-formation rate, or redshift. Such studies may be possible within the next decade using new facilities such as \textit{SOFIA} and \textit{JWST-MIRI}. 
\section{SUMMARY}\label{sum}

In summary, we find clear evidence of 10~$\micron$ and 18~$\micron$ silicate absorption in the TXS 0218+357 QAS. This shape and peak wavelength of these absorption features
appears to be best represented by an amorphous olivine template, with a measured peak optical depth of $0.46\lesssim \tau_{10} \lesssim 0.49$, although we cannot rule out the possibility of substructure 
originating in crystalline silicates in the relatively poor S/N data. A quantitative assessment of the silicate crystallinity in this system will require higher S/N data
and detection of longer-wavelength mid-IR crystalline resonance features.
Fits using amorphous pyroxene and SiC templates poorly matched the observed silicate absorption profile for the TXS 0218+357 QAS.

The derived optical depth is significantly higher than that found for other QASs, although the rest-frame 
Mg~II equivalent width and the optical depth of the 21-cm absorption are within the range of values observed for the other program QASs. This may suggest that the silicate dust in the TXS 0218+357 system
is not accompanied by a substantial enhancement in gas. We also find that while the reddening in the TXS 0218+357 system is higher than in other QASs, it is not as high
as the large optical depth would predict, given an extrapolation between these two parameters based on Milky Way diffuse clouds. This may suggest variations in the silicate dust grain 
properties between different galaxies, which can be investigated further with larger samples. 
\acknowledgments
We thank the anonymous referee for his/her comments, and Dr. Eli Dwek (NASA-Goddard) for conversations which improved the content of this paper. 
This research has made use of the NASA/IPAC Extragalactic Database (NED) which is operated by the Jet Propulsion Laboratory, California Institute of Technology, under contract with the National Aeronautics and Space Administration (NASA).
This work is based on observations made with the Spitzer Space Telescope, which is operated by the Jet Propulsion Laboratory, California Institute of Technology under a contract with NASA. 
Support for this work was provided by NASA through an award issued by JPL/Caltech. 
MCA and VPK also gratefully acknowledge partial support from the NSF grant AST-1108830 (PI Kulkarni) to the University of South Carolina.

{\it Facilities:} \facility{Spitzer (IRS)}

\appendix
\section{QUASAR CONTINUUM NORMALIZATIONS}\label{normAP}
In order to determine the effect on our conclusions of variations in the adopted underlying quasar continuum shape, we explore in this Appendix alternative quasar continuum normalizations. 

\subsection{Quasar Continuum Fitting}\label{normfitit}
The third order Chebyshev polynomial fit we adopt in our analysis was selected because it visually provided the best match to the quasar continuum shape. We
also explored 22 other fits to the continuum, produced by linear, power law, third order Chebyshev polynomial, and cubic spline fits. We found that all of the fits were similar near
10~$\micron$, while there were variations in the more poorly constrained 18~$\micron$ region because of the paucity of data at $\lambda>$21~$\micron$. We summarize in Table~\ref{contin}
the two most dissimilar fits over the silicate features. 

\begin{deluxetable}{lllll}
\tabletypesize{\scriptsize}
\tablecaption{TXS 0218+357 Quasar Continuum Variations}
\tablewidth{0pt}
\tablehead{
\colhead{Name} & \colhead{Function} & \colhead{Excl. Regions} & \colhead{Rejected Pts.} &  \colhead{Comment}}
\startdata
fiducial & 3rd Order Chebyshev & 8.5-12.5; 16.5-20.4 & LL1/LL2 join & adopted in main body of paper\\
ALTNORM1 & 3rd Order Chebyshev & 8.5-12.5; 16.5-20.4 & LL1/LL2 join; $\lambda$ near 21~$\micron$ & shallower 18~$\micron$ feature\\
ALTNORM2 & Linear & 8.5-12.5; 16.5-20.4 & \nodata & deeper 18~$\micron$ feature\\
\enddata
\tablecomments{Summary of the alternative quasar continuum normalizations. For each (column 1) normalization, we list the (column 2) fitting function,
(columns 3-4) regions and points excluded in the fitting, and (column 5) description of impact on 18~$\micron$ silicate absorption.} \label{contin}
\end{deluxetable}

\begin{figure}
\epsscale{0.7}
\plotone{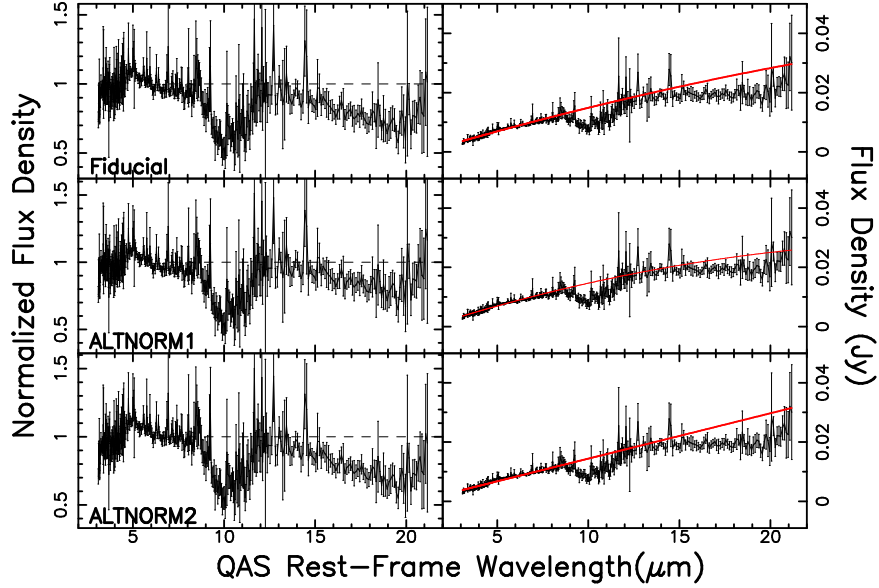}%
\caption{\textit{Left:} Normalized spectrum obtained using fiducial and 2 alternate quasar continuum fits to illustrate sensitivity of 18~$\micron$ feature on adopted normalization. 
\textit{Right:} Applied quasar continuum fits (shown in red) over-plotted on the TXS 0218+357 quasar spectrum.
[See the electronic edition of the Journal for a color version of this figure.]}\label{NORM-fig3}
\end{figure}

The two quasar continuum normalizations with the largest differences in the shapes of the silicate absorption features were obtained using the Chebyshev polynomial and linear continuum fits illustrated
in Figure~\ref{NORM-fig3}. The first normalization (ALTNORM1) uses a third order Chebyshev polynomial like the fiducial fit, but rejects several additional points at $\lambda\gtrsim$20~$\micron$ which
produces a shallower and slightly narrower 18~$\micron$ absorption feature. The second normalization (ALTNORM2) is a weighted linear fit, excluding the regions associated 
with the 10~$\micron$~silicate feature (8.5-12.5~$\micron$) and the 18~$\micron$ feature (16.5-20.4~$\micron$), and results in a slightly deeper and broader 18~$\micron$ feature. 
The differences in the 10~$\micron$~region are minimal for both normalizations. 
The shallower 18~$\micron$ feature would have a rest-frame equivalent width of 0.85$\pm$0.26 (a 3.3$\sigma$ detection), while the the broader feature would have a larger rest-frame equivalent-width of 1.53$\pm$0.35 (a 4.4$\sigma$ detection).
As previously discussed, if the feature extends redward of our data, these are lower limits.

\subsection{Silicate Absorption Feature Fits}\label{APB-fits}
\begin{figure}
\epsscale{0.8}
\plotone{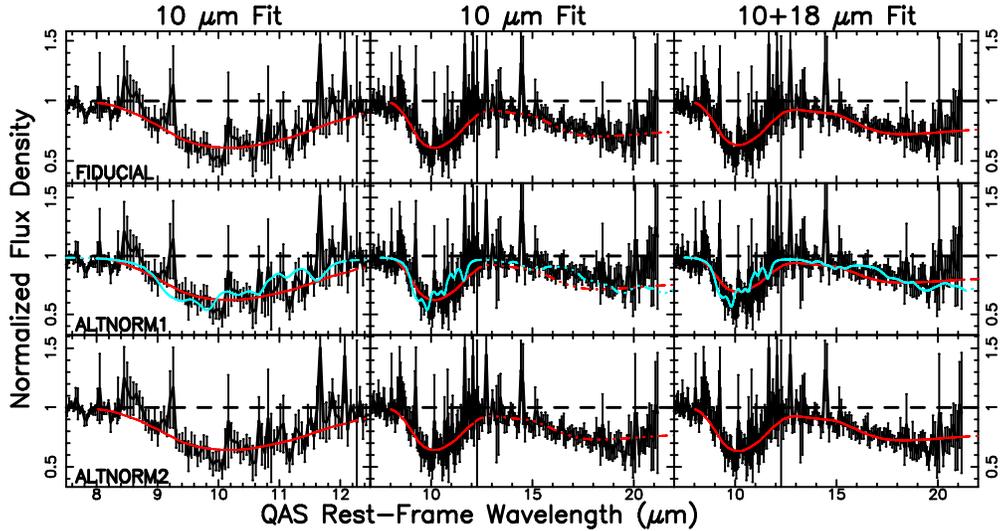}%
\caption{Similar to Figure~\ref{BESTSINGLEFIT}, but illustrating the best fits for the alternative quasar continuum normalizations. The amorphous olivine fit is shown in red, in every panel. 
For ALTNORM1 the crystalline pyroxene fit (best for the 10 and 18~$\micron$ combined fitting) is shown in cyan.} \label{NORM-fig4}
\end{figure}

For these 2 alternative quasar continuum normalizations, we repeated the fitting described in \S\ref{results}. We did not alter the $\chi^2$ minimization region in the fitting, although the
breadth of the features vary slightly between the continuum normalizations. We found that when fitting over the 10~$\micron$ region alone the alternative quasar continuum normalizations had
a small impact on the derived peak optical depth. For both normalizations we again obtained that the Amorph.Oliv.GPC template produced the best fit. The measured peak optical depth for this fit 
was $\tau_{10}= 0.47\pm0.02$ for the ALTNORM1 normalization and $\tau_{10}= 0.44\pm0.02$ for the ALTNORM2 normalization. While these are both slightly lower than the measured $\tau_{10}=0.49\pm0.02$ 
peak optical depth obtained using the fiducial normalization, they are consistent within 1.5$\sigma$. 
We illustrate in Figure~\ref{NORM-fig4} these fits over only the 10~$\micron$ feature in the left two panels, with the Am.Oliv.GPC template fit shown in red.

When we extended the fitting region to cover both the 10~$\micron$ and 18~$\micron$ features, the differences in the quasar continuum normalization were more significant. The 
shallower 18~$\micron$ feature in the ALTNORM1 normalization results in a 10:18~$\micron$ optical depth ratio that is more consistent with considered crystalline pyroxene templates than with those for amorphous olivine.
Furthermore, using the best-fitting minerals, the derived peak optical depth for the 10~$\micron$ feature was found to be deeper than derived for the fiducial fit: $\tau_{10}=0.57^{+0.03}_{-0.02}$ versus the fiducial normalization fit of $\tau_{10}=0.46\pm0.02$, as shown in Table~\ref{FITS-10+18-APB}. 
We illustrate this crystalline pyroxene fit in cyan in Figure~\ref{NORM-fig4}.
While more substructure is associated with the crystalline pyroxene fit, it is not inconsistent with our data given the relatively low S/N. 
For the ALTNORM2 normalization the peak optical depth and best-fitting mineral were consistent with the fiducial normalization. 

\begin{deluxetable}{lllllll}
\tabletypesize{\scriptsize}
\tablecaption{Template Profile Fits to TXS 0218+357 QAS 10 \& 18~$\micron$ Features}
\tablewidth{0pt}
\tablehead{
\colhead{Category} & \colhead{Mineral} & \colhead{$\tau_{10}$} & \colhead{$\chi_r^2$} & \colhead{P} & \colhead{$\tau_{10,fid}$} & \colhead{$\chi_{r,fid}^2$} }
\startdata
\cutinhead{ALTNORM1}
Amorph. Olivine & Amorph.Oliv.GPC (1) & 0.37$\pm$0.02 & 0.91 & 79.9& 0.46$\pm$0.02 & 0.70  \\
Amorph. Pyroxene & Amorph.Pyrox.GPC (1)& 0.36$\pm$0.02&1.08&21.6&0.44$\pm$0.02 & 1.34  \\
Cryst. Olivine & SanCarl.Oliv.(2) & 0.75$\pm$0.03&1.15&8.8&0.94$\pm$0.04& 1.31 \\
Cryst. Pyroxene (Best) & Nat.Enstatite (3) &0.57$^{+0.03}_{-0.02}$&0.88&87.2&0.71$\pm$0.03 & 1.00 \\
\cutinhead{ALTNORM2}
Amorph. Olivine (Best) & Amorph.Oliv.GPC (1) & 0.45$\pm$0.02&0.71&99.9&0.46$\pm$0.02 & 0.70  \\
Amorph. Pyroxene & Amorph.Pyrox.GPC (1)& 0.43$\pm$0.02&1.80&0.0&0.44$\pm$0.02 & 1.34  \\
Cryst. Olivine & Olivine (Mg$_{1.88}$Fe$_{0.12}$SiO$_4$) (3) &0.85$\pm$0.03&1.16&7.1 &0.84$\pm$0.03& 1.27 \\
Cryst. Pyroxene & Nat.Enstatite (3)& 0.70$\pm$0.03&1.08&20.8&0.71$\pm$0.03 & 1.00 \\
\enddata
\tablecomments{For each category of mineral, we present the mineral which produced the best fit over the combined 10 and 18~$\micron$ fitting regions, using the alternative quasar continuum normalizations
We list the (column 1) category; (column 2) mineral (and reference); (column 3) measured peak optical depth at the $\sim$10~$\micron$ peak feature; (column 4) reduced chi-squared; (column 5) percentage chi-squared probability; and (columns 6-7) 
the peak optical depth and reduced chi-squared for the fiducial quasar continuum normalization fit, for reference. We do not list the SiC fits which were uniformly poor.
References: (1) \citet{Chiar06} based on data from \citet{Henning99}; (2) \citet{Koike06}; \& (3) \citet{Jaeger} with tabulated data provided by J\"{a}ger.}
 \label{FITS-10+18-APB}
\end{deluxetable}

\section{SILICATE TEMPLATE PROFILES}\label{templateAP}
In order to determine the optimal peak optical depth and place \textit{some} constraints on the mineralogy of the silicate dust grains producing the 
observed 10 and 18~$\micron$ absorption features, we considered a range of absorption profile templates from the literature. 
The full list of considered minerals is presented in Table 10 of \citet{Aller12}, where we discuss the methodology for selecting these templates.
Since our goal in this analysis is to identify plausible mineral groups, not to identify the precise mineral grain properties, we discuss only the minerals which best-fit the TXS 0218+357 QAS silicate features in each of the mineral categories:
amorphous olivines and pyroxenes, crystalline olivines and pyroxenes, and SiC sources. 
We follow conventions in the literature when discussing `amorphous' silicates. We note, however, that amorphous olivine/pyroxene physically refers to amorphous silicates of an olivine/pyroxene composition, since
all olivines/pyroxenes are intrinsically crystalline \citep{Henning10}. Furthermore, we recognize that the nature of many of the amorphous olivine/pyroxene samples in the literature have been called into question, either
because of the condensation methods which may impart impurities or because of the existence of crystalline or semi-crystalline grains in the putatively amorphous samples \citep{Speck11}. These effects
do not significantly impact our conclusions.

\end{document}